\documentclass[twoside,journal]{IEEEtran}
\usepackage{bbm}
\usepackage{tipa}
\usepackage[noend]{algpseudocode}
\usepackage{algorithmicx,algorithm}


\usepackage[dvips]{graphicx}
\usepackage{amsfonts}
\usepackage{amssymb}
\usepackage{amsmath}
\usepackage{mathrsfs}
\usepackage{verbatim}
\usepackage{subfigure}
\usepackage{balance}
\usepackage{booktabs}
\usepackage{fancybox}
\usepackage{bm}
\usepackage{pifont}
\usepackage{extarrows}
\usepackage{cite}
\usepackage{color}

\begin{document}
\bibliographystyle{IEEEtran}

\title{A Fully Bayesian Approach for Massive MIMO Unsourced Random Access
\thanks{}
\thanks{\scriptsize
The work was supported in part by the
National Natural Science Foundation of China under Grant 62171364 and 61941118. (\emph{Corresponding author: Hui-Ming
Wang.})}
\author{Jia-Cheng Jiang and Hui-Ming Wang, \emph{Senior Member}, \emph{IEEE}\hspace{0.02in}}
\thanks{\scriptsize
The authors are with the School of Information and Communications Engineering, Xi'an Jiaotong
University, Xi'an, 710049, Shaanxi, P. R. China, and also with the Ministry
of Education Key Lab for Intelligent Networks and Network Security, Xi'an, 710049, Shaanxi, P. R. China. (e-mail: j1143484496b@stu.xjtu.edu.cn; xjbswhm@gmail.com.)}
\thanks{}
\thanks{}
}
\maketitle

\begin{abstract}
In this paper, we propose a novel fully Bayesian approach for the massive multiple-input multiple-output (MIMO) massive unsourced random access (URA). The payload of each user device is coded by the sparse
regression codes (SPARCs) without redundant parity bits.
A Bayesian model is established to capture the probabilistic characteristics of the overall system. Particularly, we adopt the core idea of the model-based learning approach to establish a flexible Bayesian channel model to adapt the complex environments. Different from the traditional divide-and-conquer or pilot-based massive MIMO URA strategies, we propose a three-layer message passing (TLMP) algorithm to jointly decode all the information blocks, as well as acquire the massive MIMO channel, which adopts the core idea of the variational message passing and approximate message passing. We verify that our proposed TLMP significantly enhances the spectral efficiency compared with the state-of-the-arts baselines, and is more robust to the possible codeword collisions.
\end{abstract}
\begin{IEEEkeywords}
Unsourced random access, sparse
regression codes, fully Bayesian approach, approximate message passing, variational message passing, massive MIMO.
\end{IEEEkeywords}
\section{Introduction}
Massive machine-type communication (mMTC) is one of typical application scenarios in the fifth-generation (5G) and beyond 5G (B5G) wireless networks. The core mission of mMTC is to provide cellular connectivity to millions of low-rate machine-type devices for Internet of Things (IoT) applications \cite{ToMa2016, IoT5Gera2016}. Different from the traditional human-type communication (HTC), aiming to achieve high data rates with a large packet size, the machine-type communication (MTC) applications are typical uplink-driven with packet sizes as small as a few bits \cite{BockelmannMassive2016}. Another characteristic in mMTC is that the traffic pattern is typically sporadic with only a fraction of total user devices being active \cite{SenelGrant-Free2018, LiuMassive12018}. Such features make the grant-free access control more favorable for mMTC than the grant-based access control. The grant-free access control requires a very low control overhead without the requirements of additional control signaling exchanges to facilitate the granting of resources. At the affordable cost of increased base station (BS) complexity, the collisions resolution mechanism can be appropriated designed by taking the sporadic traffic into consideration. One typical scenario for massive connectivity is that the BS first jointly detects the active users and estimates the corresponding channels, and then the identified active users are scheduled to transmit their messages \cite{LiuMassive12018}. Such a scenario is called \emph{sourced random access}, since the BS is required to identify the active users. The core mission is to resolve the joint active detection and channel estimation (JADCE) problem and some compressed sensing (CS) technologies for JADCE have been reported \cite{jiang2021, jiang20212}.

There is another line of works in the applications of mMTC, where the active user devices wish to transmit information bit sequences to the BS in an uncoordinated fashion, and the BS is only interested in the transmitted sequences but not user identifications. Such a manner is so-called \emph{unsourced random access} (URA). All the active users employ a same codebook and the decoder in the BS outputs an unordered list of the information bits. The URA framework was introduced by Polyanskiy in \cite{Polypersp}, showing particularly relevance in the
context of the IoT, along with a random coding achievability bound for its capacity in the absence of complexity constraints. Based on that, some coding schemes, including low-density parity-check codes (LDPC) code \cite{Vem2019} and polar code \cite{Pradhan2020}, have been considered in the URA literature, intending to approach to the conceptual benchmark. Indeed, URA can be established as a CS problem in a very high dimension, which utilizes the connection between the equivalence of URA and support recovery in a high dimensional CS. However, the computational overhead of such the CS problem increases exponentially with the length of information bit sequences, and the excessive size precludes the straightforward application of existing solutions. To mitigate this issue, several works have considered a divide-and-conquer approach to split the information sequences of active user devices into several blocks. On the receiver side, a inner CS-based decoder is used to recover the each block of the information sequences. Then the information blocks correspond to one original sequence are spliced together based on a outer tree decoder \cite{Amalladinne2020, Enhance2020, 2020Vamsi, Shyianov2021}, referred to the coded CS (CCS) scheme. Besides, the authors in \cite{Fengler2019} inherited the structure of CCS and used the sparse regression codes (SPARCs), and the AMP decoder to URA was applied. SPARCs is a class of channel codes for the point-to-point AWGN channel, which can achieve rates up to Shannon capacity under maximum-likelihood (ML) decoding. Later, it was shown that SPARCs can achieve capacity under AMP decoding with either power allocation or spatial coupling \cite{Barbier2017}. Based on the SPARCs, it is analyzed in \cite{2021Fengler} that the SPARCs with  concatenated coding can also achieve a vanishing per-user error probability in the limit of large blocklength and a large number of active users at sum-rates up to the symmetric Shannon capacity, motivating the use of SPARCs in URA. The AMP decoder can be adopted to achieve both efficiency and accuracy.

The study of massive URA in multiple-input multiple-output (MIMO) systems has also drawn increasing attention. Since the MIMO provides extra dimensions for the received signals, it is expected to provide higher detection accuracy and more user access. The authors in \cite{2019ale} extends the URA into a massive MIMO case. It uses the same outer tree code as \cite{Amalladinne2020} to stitch the information sequences in all the blocks, and a non-Bayesian inner decoder is designed based on a covariance-based ML (CB-ML) algorithm. It shows that the affordable number of users can be larger as the number of the receiver antennas increase, hinting the massive MIMO can provide extra spectrum efficiency in URA. However, because of the concatenated coding, some redundant parity bits are inserted in the encoding process, which highly
reduces the overall spectral efficiency. For this issue, the work \cite{Shyianov2021} were proposed to eliminate the need of concatenated coding and better utilize the massive MIMO channel. The core idea of \cite{Shyianov2021} is that the block-wise information sequences can be stitched together by clustering the estimated channel vectors produced by the inner AMP decoder. This work leads to an important revelation in massive MIMO URA that no parity bits are required when the BS is equipped with a large number of antennas, since the channel vectors can be regarded as virtual signatures of user devices to connect the pieces of information sequences. It has shown that such a consideration achieves better spectral efficiency than the CB-ML method. However, the possible codewords collisions in each block, which means more than one device chooses the same codeword, will result in a superimposition of the estimated channel vectors, leading to the failure of the stitching process. This significantly damages the decoding performance.

Above works in the context of massive MIMO URA can be considered as the extensions of the divide-and-conquer strategy. Recently, some pilot-based approaches have been proposed in \cite{PilotFengler, PilotAhmadi, FASURA, 2022Li} to show a significantly better performance than the above divide-and-conquer approaches in the setting where the coherence length is larger than the number of the active user devices. The method in \cite{PilotFengler} is the original version of the pilot-based methods, where channel estimation is obtained in the pilot phase, and the decoding algorithm is performed in the decoding phase by considering the channel coefficients are known. The polar code associated with the list decoder has been adopted in the decoding phase. The method in \cite{PilotAhmadi} extended the work \cite{PilotFengler}, where the multiple stages of orthogonal pilots and an additional successive interference
cancellation (SIC) operation are adopted to cope with the user collisions in the pilot phase and enhance the decoding performance. The
method in \cite{FASURA} combined the pilot-based method with an additional operation called noisy pilot channel estimation (NOPICE) to re-estimate the channel based on the observed signals in both the pilot and the data phases, and further reupdate the estimations of the coded symbols.
In \cite{2022Li}, the pilot phase is encoded by means of CS coding. An approximate message passing (AMP) decoder was designed to recover information and the corresponding channel coefficients. The decoding phase is encoded by LDPC codes. Based on the channel coefficients estimated by the pilot phase, the information sequences can be obtained based on belief propagation (BP). A critical issue of these pilot-based approaches is that the decoding performance is sensitive to the channel estimation results, so the overall performance is limited by the accuracy of the pilot phase recovery. The reason behind the issue is that the recovery in the pilot phase and decoding phase are performed independently rather than jointly, which causes the channel estimation recovery in the pilot phase will not use the knowledge of the received signals in the decoding phases. This seriously affects the performance of channel estimation and thus the final decoding performance.


In this paper, we design a novel scheme to cope with all these issues in massive MIMO URA. The payload of each user device is SPARCs coded without redundant parity bits for stitching the information blocks. The SPARCs map all the blocks of information sequences for one user into one codeword, and we naturally propose a decoder to jointly recover all the blocks, acquiring both the information sequences and the massive MIMO channel.
A Bayesian model is established to capture the probabilistic characteristics of the overall system. Particularly, we adopt the core idea of the model-based learning approach to establish a flexible channel model to adapt the complex environments. A novel MP-based decoder is then designed to jointly decode all the blocks based on the Bayesian model. Our contributions can be summarized as follow.
\begin{itemize}
\item We extend the SPARCs into the massive MIMO URA. We introduce a Gaussian mixture (GM) distribution for channel, which has been demonstrated  to better capture the characteristics of the massive MIMO systems with multi-users than the Rayleigh channel \cite{CEMoG2015}. We then establish a fully Bayesian model based on the system model, channel model and coding scheme.
\item
      We design a novel decoding algorithm based on the established probabilistic model. To handle the issues that the forms of messages are too complicated, and the total number of messages is too large, we use the core idea of the variational message passing (VMP) and AMP. The whole decoding algorithm can be divided into two parts, called the \emph{channel equalization} part and the \emph{decoding} part, respectively.  In the channel equalization part, the VMP converts the complex problem into a simpler bilinear problem by minimizing the Kullback-Leibler (KL) divergence, and we therefore integrate the bilinear generalized AMP (BiG-AMP) \cite{big-amp} into this part, while the estimations of the marginal posterior distributions of channel coefficients are updated. In the decoding part, the sub-problem can be tailored into a vector-valued linear problem, and we then propose a vector-valued GAMP (Ve-GAMP) for the sub-problem, and the SPARCs are decoded based on the derived approximate posterior distributions.
    \item
    We then provide some implementation details of the proposed decoding algorithm. We modified the algorithm by the scale valued simplifications in order to further reduce its computational complexity. To reduce the chance of the misconvergence of our algorithm, we damp the iterations by some damping factors. Besides, we provide a cost function to measure the similarity between the estimated marginal posterior distributions obtained by our algorithm and the exact one. This is done by using the definition of KL divergence. The cost function can be used in each iteration to evaluate the algorithm performance, monitor the convergence and adjust the damping factors.

\end{itemize}
The rest of the paper is organized as follows. In Section II, we introduce the system model and probabilistic model for the MIMO massive URA. In Section III, we propose a novel Bayesian decoding algorithm based on AMP and VMP. In Section IV, some implementation details for simplifying the decoding algorithm, avoiding misconvergence and evaluating the algorithm performance are provided. After verifying the numerical results and in Section V, we conclude the paper in Section VI.

\section{System Model}
Consider the uplink of a single-cell cellular network, where a central $M$-antennas base station (BS) serves $K_{tot}$ single-antenna devices in total. This paper considers the sporadic device activity, where only a relatively small number $K$ of devices are active within one coherence time. We note that the number of the active users is known for the BS. We assume the active devices synchronously transmit a message loading $B$ bits of information into a block-fading channel with $N$ channel uses, where the channel coefficients are considered unchanged during the $N$ channel uses. Let $\tilde{\mathbf b}_k\in \{0,1\}^B$ denote the binary information sequence of the device $k$, $\mathbf r_k\in \mathcal C\subset \mathbb R^N$ the corresponding codeword taken from a common codebook $\mathcal C$ and $f(\cdot): \{0,1\}^B\rightarrow \mathbb R^N$ the encoding function that maps the information sequence $\tilde{\mathbf b}_k$ to the codeword $\mathbf r_k$. The received signal can be formulated as
\begin{align}
\tilde{\mathbf Y} = \sum\nolimits_{k = 1}^K f(\tilde{\mathbf b}_k)\tilde{\mathbf h}_k^T+\tilde{\mathbf W},\label{sig}
\end{align}
where $\tilde{\mathbf h}_k\in \mathbb R^{M\times 1}$ is the channel vector of device $k$ and $\tilde{\mathbf W}\in \mathbb R^{N\times M}$ is the additive white Gaussian noise (AWGN) matrix with each element distributed as $\mathcal{N}(0,\sigma^2)$. Note that the parameters in the signal model (\ref{sig}) are all considered in the real domain. We would like to mention that the more realistic signal model assuming the complex channel can be easily transformed into (\ref{sig}), the subsequent derived algorithm can be easily transformed into its complex version, and we use (\ref{sig}) for the ease of following derivations. The task of the decoder in the BS is to produce a list $g(\tilde{\mathbf Y})$ consisting of the information sequences of the $K$ active devices. We note that the BS is not required to identify where these sequences are sent, thereby leading to the so-called URA. To examine the performance of a decoder in the URA system, we introduce the per-user probability \cite{Amalladinne2020, Shyianov2021, 2021Fengler}, where an error is declared if one of the transmitted messages is not in the output list  $g(\tilde{\mathbf Y})$. The probability can be written as
\begin{align}
P_e = K^{-1}\sum\nolimits_{k = 1}^K {\rm Pr}\big(\tilde{\mathbf b}_k\notin g(\tilde{\mathbf Y})\big).\label{pe}
\end{align}
\subsection{GM Channel Model}
We here establish the probabilistic model over the massive MIMO channel. The Rayleigh channel model is used in the existing benchmarks \cite{2019ale, Shyianov2021, 2022Li} that combines the MIMO and URA systems. However, such a channel model assumption is not sufficiently realistic, since elements in each $\mathbf h_k$ have different variances and exhibit a high level of sparsity due to the limited angular spread. To handle this issue, we in this paper adopt the core idea of model-based learning approach \cite{JW2021}. To avoid the performance loss caused by the inaccurate channel model, we establish a model set, consisting of a broad class of candidate models, and the model is then refined by learning the model parameters. To this end, we introduce the GM distribution for channel, which has been demonstrated  to better capture the characteristics of the massive MIMO systems with multi-users than the Rayleigh channel \cite{CEMoG2015}.

In a typical cellular communications system, the channel from a device to a BS is spatially correlated with a covariance matrix depending on the realistic scattering geometry. By adopting an $M\times M$ matrix consisting of a set of orthogonal bases, denoted as $\mathbf F$, the channel vector for the $k$th device can be transformed into a virtual channel \cite{JSDM2013} represented as $\mathbf h_k = \mathbf F^T\tilde{\mathbf h}_k$. One of the most crucial property of the virtual channel vector is that the elements are independent and approximated sparse, i.e., only a small fraction of elements is in a large scale, and others are approximately zero. We therefore adopt the GM distribution to model the virtual channel, capturing the significantly different variances of each channel coefficient. Specifically, the probability density function (PDF) over each element $h_{km}$ in the channel vector $\mathbf h_k$ is
\begin{align}
p(h_{km}) = \sum\nolimits_{g = 1}^G \pi_g \mathcal{N}(h_{km};0,\gamma_g^{-1}),\label{Gmchannel}
\end{align}
where $\pi_g$ is the mixing coefficient of all the components that satisfies $\sum\nolimits_{g = 1}^G \pi_{g} = 1$, and $\gamma_g = \sigma_g^{-2}$ is precision of $g$th component. The GM distribution is flexible and controlled by the parameters $(\pi_1|\dots|\pi_G)$, $(\gamma_1|\dots|\gamma_G)$ and $G$, which are determined by the parameters in the realistic scenarios, i.e., the large-scale fading coefficients, angular spreads and angles of arrival of devices.

\subsection{Encoding}
In this paper, we focus on the so-called  sparse regression codes (SPARCs). Such a scheme has been proved to achieve a vanishing error probability in an asymptotic regime at sum-rates up to the Shannon capacity in both the single-user system \cite{AMPdecoder} and the multi-user single-antenna URA system \cite{2021Fengler}. We thereby aim to extend the works to the MIMO URA system.

The transmitted codeword is established in the following way. The $B$-bit message $\tilde{\mathbf b}_k$ is divided into $J$ equal-length sub-sequences with length $L = B/J$. For each block of the $L$-bit sub-sequence, we define a variable $\mathbf x_{jk}\in\{0,1\}^{1\times 2^L}$ with only one non-zero element, for $j = [1:J]$ and $k = [1:K]$. The common codebook $\mathcal C$ is based on a set of $J$ coding matrices $\mathbf A_j = (\mathbf a_{1j}|\dots|\mathbf a_{Nj})^T\in \mathbb R^{N\times 2^L}$, which is known for both the BS decoder and devices.  The codeword for device $k$ is created by
\begin{align}
\mathbf r_{k} = f(\tilde{\mathbf b}_k) = \sum\nolimits_{j = 1}^J\mathbf A_j\mathbf x_{jk} = \mathbf A\mathbf x_k, \label{cod}
\end{align}
where $\mathbf A = (\mathbf A_1|\dots|\mathbf A_J)$ and $\mathbf x_k = (\mathbf x_{1k}|\dots|\mathbf x_{Jk})^T$. Each element in matrix $\mathbf A$ is distributed as $\mathcal{N}(0, \frac{1}{NJ})$ to satisfy the unit power constraint of codewords. It is typical to assume the power coefficients for devices to be a constant that is absorbed by the noise variance, denoted as $\sigma^2$.

\subsection{Fully Bayesian Model}
Here, we establish a fully Bayesian model based on the system model (\ref{sig}), channel model (\ref{Gmchannel}) and coding scheme (\ref{cod}), and the Bayesian inference can be executed based on this model. We can infer from (\ref{cod}) that the system model (\ref{sig}) can be rewritten in a matrix form as
\begin{equation}
\mathbf Y = \mathbf A\mathbf X\mathbf H + \mathbf W,\label{con_mod}
\end{equation}
where $\mathbf Y= \tilde{\mathbf Y}\mathbf F$ is $N\times M$, $\mathbf X = (\mathbf x_1|\dots|\mathbf x_K)$ is $2^LJ\times K$, and $\mathbf H = (\mathbf h_1|\dots|\mathbf h_K)^T$ is $K\times M$, remaining constant for the whole frame. The $\mathbf W = \tilde{\mathbf W}\mathbf F$ is still an AWGN matrix with a variance $\sigma^2$, since in the typical consideration, the $\mathbf F$ is an orthogonal matrix. Accordingly, the measurement $\mathbf Y$ can be specified by a probability distribution function $p(\mathbf Y|\mathbf R, \mathbf H)$ on the condition of $\mathbf R = (\mathbf r_1|\dots|\mathbf r_K)$ and $\mathbf H$, which is given by
\begin{align}
p(\mathbf Y|\mathbf R, \mathbf H) &= \prod_{nm}p(y_{nm}|\sum\limits_{k}r_{nk}h_{km}\nonumber \\
&= \prod_{nm}\mathcal{N}(y_{nm};\sum\limits_{k}r_{nk}h_{km}, \sigma^2),\label{like}
\end{align}
where $y_{nm}$, $r_{nk}$ and $h_{km}$ are the elements in the matrix $\mathbf Y$, $\mathbf R$ and $\mathbf H$, respectively. The coding scheme in (\ref{cod}) then immediately yields
\begin{align}
p(\mathbf R|\mathbf X) = \prod_{n,k} \delta(r_{nk} = \sum\limits_{j}\mathbf a_{nj}^T\mathbf x_{jk}), \label{rconx}
\end{align}
where $\delta(\cdot)$ is the Dirac delta function, We note that we have no additional restrictions on the sequences of bits sent by devices, thereby the prior distribution for the $j$th sub-sequence of $k$th device can be typically designed to give uniform weight to any permutation of the $2^L$-dimensional vector $\mathbf x_{jk}$, given by
\begin{align}
p(\mathbf x_{jk}) = \frac{1}{2^L}\sum\nolimits_{i = 1}^{2^L}\delta(x_{jk,i}-1)\prod_{m\neq i}\delta(x_{jk,m}).\label{pri_x}
\end{align}
For the prior distribution (\ref{Gmchannel}) of the channel vector, we introduce a series of latent variable, denoted as $\boldsymbol\Omega$, to facilitate the subsequent parameters learning process.
 For each element $h_{km}$, there is a corresponding latent variable $\boldsymbol\omega_{km} = (\omega_{km,1}|\dots|\omega_{km,G})$, satisfying $\sum\limits_{g}\omega_{km,g} = 1$ and $\omega_{km,g}\in \{0,1\}$. This means that each $\boldsymbol\omega_{km}$ has a multinomial
distribution. With the given latent variable, the PDF of the element in the channel vector can be re-written as
\begin{align}
p(h_{km}|\boldsymbol\omega_{km}) = \prod_{g}\mathcal{N}(h_{km};0,\gamma_g^{-1})^{\omega_{km,g}},\label{h_cono}
\end{align}
with the prior distribution of the latent variable
\begin{align}
p(\boldsymbol\omega_{km}|\boldsymbol\beta) = \prod_{g}\pi_g^{\omega_{km,g}} = \prod_{g}\big(\beta_g\prod_{t = 1}^{g-1}(1-\beta_t)\big)^{\omega_{km,g}}.\label{wconb}
\end{align}
We note that we have introduced a series of the parameters $\boldsymbol\beta = \{\beta_1,\dots, \beta_G\}$ with each parameter $\beta_g$ distributed a beta distribution, given by
\begin{align}
p(\boldsymbol\beta) = \prod_{g}B(1,\alpha)^{-1}(1-\beta_g)^{\alpha-1},\label{pri_be}
\end{align}
where $B(1,\alpha)$ is the normalization factor, and $\alpha$ is a small constant. The connection between the mixing coefficient $\pi_g$ and $\boldsymbol\beta$ can be specified as
$
\pi_g = \beta_g\prod_{t = 1}^{g-1}(1-\beta_t)
$.
According to \cite{JW2021}, by introducing $\boldsymbol\beta$, the mixing coefficient can be regarded as generated from a Dirichlet process (DP), thereby the so-called DP prior distribution for the mixing coefficients. Such a prior distribution offers full flexibility in deciding the number of GM channel components, automatically determines the amount of components $G$ that is required to model the realistic channel distribution without a parameters tuning process. For consistency, we also introduce the prior distribution ${\rm Gam}(\boldsymbol\gamma; a, b)$ over parameters $\boldsymbol\gamma = (\gamma_1|\dots|\gamma_G)$ in (\ref{h_cono}), given by
\begin{align}
p(\boldsymbol\gamma) = \prod_g \Gamma(a,b)^{-1}\gamma_g^{a-1}\exp(-b\gamma_g),\label{pri_g}
\end{align}
where $\Gamma(a,b)$ is a normalization factor and $a$, $b$ are some small constants. We are then capable to establish the fully Bayesian model over (\ref{con_mod}). The joint distribution of all of the random variables can be written down by
\begin{align}
&p(\mathbf Y, \boldsymbol\Theta) = \prod_{n,m}p\big(y_{n,m}|\sum\limits_{k}r_{nk}h_{km}\big)\prod_{n,k}\delta\big(r_{nk} = \sum\limits_{j}\mathbf a_{nj}^T\mathbf x_{jk}\big)
\nonumber\\
\times&\prod_{k,m}p(h_{km}|\boldsymbol\omega_{km},\boldsymbol\gamma)p(\boldsymbol\omega_{km}|\boldsymbol\beta)p(\boldsymbol\gamma)p(\boldsymbol\beta)\prod_{j,k}p(\mathbf x_{jk}),\label{jot_dis}
\end{align}
where $\boldsymbol\Theta=\{\mathbf X, \mathbf R, \mathbf H, \boldsymbol\Omega, \boldsymbol\gamma, \boldsymbol\beta\}$, and the evolved factors are specified in
(\ref{like})-(\ref{pri_g}), respectively. These forms of prior distributions over $\mathbf H, \boldsymbol\Omega, \boldsymbol\gamma, \boldsymbol\beta$ are conjugate prior distributions to permit the subsequent approximate posterior distributions in the same functional forms of their prior distributions. Such a Bayesian treatment has some critical distinctions compared with the existing works \cite{2019ale, Shyianov2021, 2022Li} in the MIMO URA systems. We do not require the redundant parity bits to be inserted aiming to stitch all the sequences together. We establish a fully Bayesian model for all the blocks, instead of considering each block as an independent CS instance. Such a consideration leads to that the observations in each block will benefit the recovery of all other blocks. As we shall see, executing the Bayesian inference on (\ref{jot_dis}), our decoding algorithm will achieve significant spectral efficiency gain and have much stronger robustness to the possible codewords collisions compared with its counterparts.

The task of the decoder then turns to produce the output list of information sequences $g(\tilde{\mathbf Y})$ based on the joint distribution (\ref{jot_dis}) to minimize the per-user probability (\ref{pe}). One way to reach this goal is to design a decoder based on the minimum mean square error (MMSE) criterion, which requires to produce the mean of the posterior distribution based on our probabilistic model. However, the exact posterior distribution for each $\mathbf x_{jk}$ is unfeasible to compute based on (\ref{jot_dis}), we thereby design an approximated decoding algorithm to achieve sub-Bayes-optimality.

\section{Novel Bayesian Decoding Algorithm}\label{de-al}
We in this section design a decoding algorithm to execute the Bayesian inference on (\ref{jot_dis}).  For the ease of conciseness, we artificially divide the whole decoding algorithm into two parts, called the \emph{channel equalization} part and the \emph{decoding} part, respectively. In the channel equalization part, the estimations of the marginal posterior distributions of channel coefficients are updated. In the decoding part, the SPARCs are decoded based on the derived approximate posterior distributions. The two parts of the decoding algorithm are jointly updated. Our decoding algorithm lies on the basic theory of MP algorithms, where the beliefs are propagated along with the messages \cite{Bishop2006Pattern}. However, the standard BP and loopy BP, due to their low efficiency in computation, are unfeasible to be executed in our scenario. The low efficiency comes from two aspects, i.e., the forms of messages are too complicated, and the total number of messages is too large. We therefore adopt the core idea of VMP \cite{2005Variational} and AMP \cite{DonohoMessage} to deal with such issues. We note that the Bayesian inference on our model is different from that in the typical linear mixing \cite{GAMP} or the bilinear mixing problems \cite{big-amp} in the URA systems, since the received signals $\mathbf Y$ are obtained by the mixing of three matrices, i.e., $\mathbf A$, $\mathbf X$ and $\mathbf H$, and our proposed decoding algorithm is so-called three-layer MP (TLMP) algorithm.

\subsection{Part I: Channel Equalization}
\label{kl}
\subsubsection{Approximate the channel prior distribution via VMP}
We can see in (\ref{h_cono})-(\ref{pri_g}), the complex forms of the prior distributions over random variables $\mathbf H$, $\boldsymbol\Omega$, $\boldsymbol\beta$ and $\boldsymbol\gamma$ make the execution of BP algorithm in the channel equalization part too complicated. We therefore adopt the VMP framework at the beginning of the channel equalization part, which aims to convert such a complex problem into a simpler problem by using an approximating distribution that has a simpler dependency structure than that of the exact solution. Typically, VMP uses a distribution in a factorized form to approximate the true posterior distribution with each factor associated with a particular subset of random variables in $\boldsymbol\Theta$. This assumption neglects the dependency of $\mathbf H$, $\boldsymbol\Omega$, $\boldsymbol\beta$, and $\boldsymbol\gamma$ conditioned on $\mathbf Y$, which results in a simpler probabilistic structure, as we will see in the following. To this end, we partition random variables in $\boldsymbol\Theta$ into different subsets, and introduce a series of distributions, using notation $q(\cdot)$, to express the approximations of their posterior distributions. Accordingly, we can approximate $p(\boldsymbol\Theta|\mathbf Y)$ by
\begin{align}
p(\boldsymbol\Theta|\mathbf Y) \approx q(\mathbf H, \mathbf R, \mathbf X)q(\boldsymbol\Omega)q(\boldsymbol\beta)q(\boldsymbol\gamma),\label{app_pos}
\end{align}
Note that the random variables $\mathbf X$, $\mathbf R$ and $\mathbf H$ are not be partitioned into different factors. As we will see later in this section, the approximated posterior distribution $q(\mathbf H, \mathbf R, \mathbf X)$ provides an appropriate form to adopt the AMP-like approximation, resulting in that it is not necessary to decouple the dependency between these random variables. VMP then aims to minimize the similarity between the approximated posterior in (\ref{app_pos}) and the true one. For this purpose, the KL divergence is used to measure such a similarity, which can be formulated as
$
\mathcal{D}_{{\rm KL}}[q(\boldsymbol\Theta)||p(\boldsymbol\Theta|\mathbf Y)]  = -\mathcal{L}[q(\boldsymbol\Theta)]+p(\mathbf Y)
$,
where $\mathcal{L}[q(\boldsymbol\Theta)] = \langle\ln p(\boldsymbol\Theta, \mathbf Y)-\ln q(\boldsymbol\Theta)\rangle_{q(\boldsymbol\Theta)}$ and $\langle\cdot\rangle$ returns the expectation of the input random variables. To solve this optimization problem, we consider to optimize each involved factor in turn. Specifically, for a given factor $q(\tilde{\boldsymbol\theta})$, by invoking the results of the VMP methodology \cite{2005Variational}, we can easily deduce that the optimal $q(\tilde{\boldsymbol\theta})$ that maximizes $\mathcal{L}[q(\boldsymbol\Theta)]$ with given other factors satisfies
\begin{equation}
\ln q(\tilde{\boldsymbol\theta}) = \langle\ln p(\mathbf Y, \boldsymbol\Theta)\rangle_{q(\boldsymbol\Theta)\setminus q(\tilde{\boldsymbol\theta})}+{\rm const},\label{log_opt}
\end{equation}
where $\langle\cdot\rangle_{q(\cdot)}$ denotes the expectation with respect to $q(\cdot)$, $q(\boldsymbol\Theta)\setminus q(\tilde{\boldsymbol\theta})$ denotes the factorized distribution $q(\boldsymbol\Theta)$ without the factor $q(\tilde{\boldsymbol\theta})$, and the notation '${\rm const}$' represents a constant that does not depend on $\tilde{\boldsymbol\theta}$. We note that each factor is evaluated by using the current estimates for all the others, so that (\ref{log_opt}) should be executed alternatively.
We then provide the detailed derivations of the involved factors $q(\mathbf H, \mathbf R, \mathbf X)$, $q(\boldsymbol\Omega)$, $q(\boldsymbol\beta)$ and $q(\boldsymbol\gamma)$ based on the general VMP principle in (\ref{log_opt}).

\emph{For the factor $q(\boldsymbol\Omega)$}, we have
\begin{align}
\ln &q(\boldsymbol\Omega)\nonumber\\
&= \langle\ln p(\mathbf H|\boldsymbol\Omega, \boldsymbol\gamma)\rangle_{q(\mathbf H,\mathbf R,\mathbf X)q(\boldsymbol\gamma)}+\langle\ln p(\boldsymbol\Omega|\boldsymbol\beta)\rangle_{q(\boldsymbol\beta)}+{\rm const}\nonumber\\
&= \sum\nolimits_{k,m,g} \omega_{km,g}\bigg(\frac{1}{2}\langle\ln\gamma_{g}\rangle-\frac{1}{2}\langle\gamma_{g}\rangle
 \langle|h_{k,m}|^2\rangle+\langle\ln\beta_g\rangle\nonumber\\
&\quad\quad+\sum\nolimits_{p = 1}^{g-1}\langle\ln(1-\beta_p)\rangle\bigg)+{\rm const}.\label{qome}
\end{align}
The obtained factor $q(\boldsymbol\Omega) = \prod_{km} q(\boldsymbol\omega_{km})$ in (\ref{qome}) is in the same functional form of its corresponding prior distribution (\ref{wconb}) with parameters $\boldsymbol\zeta_{km} = (\zeta_{km,1}|\dots|\zeta_{km,G})$ for each $k$ and $m$,
 where we let \begin{align}
\ln\zeta_{km,g} = \frac{1}{2}\langle\ln\gamma_{g}\rangle&-\frac{1}{2}\langle\gamma_{g}\rangle
 \langle|h_{k,m}|^2\rangle+\langle\ln\beta_g\rangle\nonumber\\
 &+\sum\nolimits_{p = 1}^{g-1}\langle\ln(1-\beta_p)\rangle.\label{lnzeta}
 \end{align}
Another critical observation is that the received beliefs from the involved nodes $\mathbf H$, $\boldsymbol\gamma$ and $\boldsymbol\beta$ are reflected on their statistics, i.e.,
$\langle\ln\gamma_{g,m}\rangle$, $\langle\gamma_{g,m}\rangle$, $\langle|h_{k,m}|^2\rangle$, $\langle\ln\beta_g\rangle$ and $\langle\ln(1-\beta_g)\rangle$, which will be derived later. As for $\boldsymbol\Omega$, the useful expectation for each element
$\omega_{km,g}$ can be formulated as
\begin{equation}
\langle\omega_{km,g}\rangle = \zeta_{km,g}\left(\sum\nolimits_{g}\zeta_{km,g}\right)^{-1}.\label{zeta}
\end{equation}
\emph{For the factor $q(\boldsymbol\beta)$}, we have
\begin{align}
\ln q(\boldsymbol\beta)
 = &\sum\nolimits_{k,m}\langle\ln p(\boldsymbol\omega_{km}|\boldsymbol\beta)\rangle_{q(\boldsymbol\Omega)}+\ln p(\boldsymbol\beta)+{\rm const}\nonumber\\
= &\sum\nolimits_{k,m}\sum\nolimits_{g}\langle \omega_{km,g}\rangle\big(\ln \beta_g+\sum\nolimits_{t = 1}^{g-1}\ln(1-\beta_t)\big)\nonumber\\
&\quad\quad+\sum\nolimits_{g}(\alpha-1)\ln(1-\beta_g)+{\rm const}\nonumber\\
 = &\sum\nolimits_{g}\big(\sum\nolimits_{k,m}\sum\nolimits_{p =g+1}^G\langle \omega_{km,p}\rangle+\alpha-1\big)\ln(1-\beta_g)\nonumber\\
 &\quad\quad+\sum\nolimits_{k,m,g}\langle \omega_{km,g}\rangle\ln\beta_g+{\rm const}.\label{qbet}
\end{align}
We note that the obtained $q(\boldsymbol\beta) = \prod_{g}q(\beta_g)$ can be factorized, and each element $\beta_g$ is distributed as a beta distribution $\beta_g\sim\mathcal{B}(\tau_g,\tilde{\tau}_g)$ with parameters $\tau_g$ and $\tilde{\tau}_g$, given by
\begin{align}
\tau_g &= \sum\nolimits_{k,m}\langle \omega_{km,g}\rangle+1, \nonumber\\
\tilde{\tau}_g &= \sum\nolimits_{k,m}\sum\nolimits_{p =g+1}^G\langle \omega_{km,p}\rangle+\alpha.\label{tau}
\end{align}
Based on that, the involved expectations of logarithmic forms are specified as
\begin{align}
\langle\ln\beta_g\rangle &= \psi\left(\tau_g\right)-\psi\left(\tau_g +\tilde{\tau}_g\right),\nonumber\\
\langle\ln(1-\beta_g)\rangle &= \psi\left(\tilde{\tau}_g)-\psi(\tau_g +\tilde{\tau}_g\right),
\end{align}
where $\psi(x) = \frac{{\rm d}\ln\Gamma(x)}{{\rm d}x}$ is the digamma function. Then, for the factor $q(\boldsymbol\gamma)$, we have
\begin{align}
\ln q(\boldsymbol\gamma)
&= \sum\nolimits_{k,m}\langle\ln p(h_{km}|\boldsymbol \omega_{km}, \boldsymbol\gamma)\rangle_{q(\mathbf H,\mathbf R, \mathbf X)q(\mathbf Z)}\nonumber\\
&\quad\quad+\sum\nolimits_{g}\ln p(\boldsymbol\gamma)+{\rm const}\nonumber\\
&= \sum\nolimits_{k,m}\sum\nolimits_{g}\langle \omega_{km,g}\rangle\left(\frac{1}{2}\ln\gamma_{g}-\frac{1}{2}\gamma_{g}\langle|h_{k,m}|^2\rangle\right)\nonumber\\
&\quad\quad+\sum\nolimits_{g}(a-1)\ln\gamma_{g}-b\gamma_{g}+{\rm const}\nonumber\\
& = \sum\nolimits_{g}\sum\nolimits_{k,m}\left(\frac{1}{2}\langle \omega_{km,g}\rangle+a-1\right)\ln\gamma_{g}\nonumber\\
&\quad\quad-\left(\frac{1}{2}\langle \omega_{km,g}\rangle\langle|h_{k,m}|^2\rangle+b\right)\gamma_g+{\rm const}.\label{qgam}
\end{align}
We note that the obtained $q(\boldsymbol\gamma) = \prod_{g}q(\gamma_g)$ can be factorized with each element independently distributed as $q(\gamma_{g,m}) \sim {\rm Gam}(\gamma_{g};\tilde{a}_{g},\tilde{b}_{g})$. The corresponding parameters are
\begin{align}
\tilde{a}_{g} &= \sum\nolimits_{k,m}\frac{1}{2}\langle \omega_{km,g}\rangle+a, \nonumber\\
\tilde{b}_{g} &= \sum\nolimits_{k,m}\frac{1}{2}\langle \omega_{km,g}\rangle\langle|h_{k,m}|^2\rangle+b.\label{para_gamma}
\end{align}
As a consequence, we have the expectations
$
\langle\gamma_{g}\rangle = \tilde{a}_{g}\big(\tilde{b}_{g}\big)^{-1}$ and $\langle\ln\gamma_{g}\rangle = \psi(\tilde{a}_{g})-\ln\tilde{b}_{g}
$.

\emph{For the factor $q(\mathbf H, \mathbf R, \mathbf X)$}, we have
\begin{align}
\ln& q(\mathbf X, \mathbf R, \mathbf H)\nonumber\\
= &\sum\nolimits_{n,m}\ln p_{{\rm out}}(y_{n,m}|\sum\nolimits_{k}r_{nk}h_{km})\nonumber\\
&\quad\quad+\sum\nolimits_{n,k}\ln \delta(r_{nk} = \sum\nolimits_{j}\mathbf a_{nj}^T\mathbf x_{jk})+\sum\nolimits_{j,k}\ln p(\mathbf x_{jk})\nonumber\\
+ &\sum\nolimits_{k,m}\langle\ln p(h_{km}|\boldsymbol\omega_{km},\boldsymbol\gamma)\rangle_{q(\boldsymbol\Omega)q(\boldsymbol\gamma)}.\label{ln_qh}
\end{align}
The expectation term in the final equation can be specified as
\begin{align}
&\sum\nolimits_{k,m}\langle\ln p(h_{km}|\boldsymbol\omega_{km},\gamma_g)\rangle_{q(\boldsymbol\Omega)q(\boldsymbol\gamma)}\nonumber\\
 = &\sum\nolimits_{k,m}\sum\nolimits_{g} \langle \omega_{km,g}\rangle\left(\frac{1}{2}\langle\ln\gamma_{g}\rangle-\langle\gamma_{g}\rangle|h_{km}|^2\right)+{\rm const}\nonumber\\
  = &\sum\nolimits_{k,m}\ln \mathcal{N}(h_{km};0, \bar{\gamma}_{km}^{-1})+{\rm const},\label{eq_prih}
\end{align}
where we define $\bar{\gamma}_{km}= \sum\nolimits_{g}\langle \omega_{km,g}\rangle\langle\gamma_{g}\rangle$. Substituting (\ref{eq_prih}) into (\ref{ln_qh}), we get the following formula
\begin{align}
q(\mathbf X, \mathbf R, \mathbf H) \propto p(\mathbf Y|\mathbf H,\mathbf R)p(\mathbf R|\mathbf X)p(\mathbf X)\tilde{p}(\mathbf H),\label{tran_mod}
\end{align}
where the distribution $\tilde{p}(\mathbf H) = \prod_{km}\tilde{p}(h_{km})$ is a Gaussian distribution, with each factor given by
\begin{align}
\tilde{p}(h_{km}) =  \mathcal{N}(h_{km};0, \bar{\gamma}_{km}^{-1}).\label{pri_h_tran}
\end{align}
We can notice from (\ref{tran_mod}) that the obtained approximated posterior distribution $q(\mathbf X, \mathbf R, \mathbf H)$ in each iteration can be regarded as the posterior distribution of a newly produced probabilistic model
\begin{align}
p(\mathbf Y, \tilde{\boldsymbol\Theta}) = p(\mathbf Y|\mathbf H,\mathbf R)p(\mathbf R|\mathbf X)p(\mathbf X)\tilde{p}(\mathbf H),\label{new_mod}
\end{align}
where we define a new set of parameters as $\tilde{\boldsymbol\Theta} = \{\mathbf X, \mathbf R, \mathbf H\}$, and the forms of the factors can be seen in (\ref{like}), (\ref{rconx}), (\ref{pri_x}) and (\ref{pri_h_tran}), respectively. The PDF in (\ref{pri_h_tran}) can be considered as the corresponding prior distribution over $\mathbf H$ of the newly obtained model (\ref{new_mod}).
In summary, VMP updates the statistics of involved random variables $\boldsymbol\Omega$, $\boldsymbol\gamma$ and $\boldsymbol\beta$ by equations (\ref{zeta}), (\ref{tau}), (\ref{para_gamma}), and convert the complex model (\ref{jot_dis}) into a simpler one (\ref{new_mod}) by using an approximating distribution (\ref{app_pos}) that has a simpler dependency structure. Note that the channel prior distribution in the transformed model is approximated by a Gaussian distribution (\ref{pri_h_tran}). According to (\ref{lnzeta}) and (\ref{para_gamma}), in order to complete the VMP, we require to obtain the marginal posterior distribution over each $h_{km}$ by the transformed model (\ref{new_mod}). This can be done by performing an AMP-like approximation, which will be discussed later.
\begin{figure}[!t]
\centering
\includegraphics[width=3in]{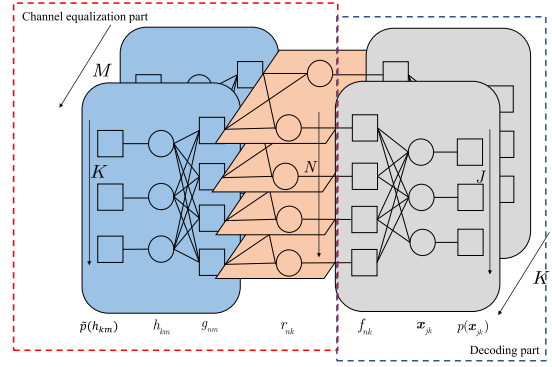}
\caption{The factor graph associated with the transformed model (\ref{new_mod}).}
\label{sys}
\end{figure}

\subsubsection{Approximate the channel posterior distribution via AMP}
The factor graph associated with the transformed model is depicted in Fig. \ref{sys}. The factor node that connects each pair of the variable nodes $h_{km}$ and $r_{nk}$ is denoted as $g_{nm}$, and the factor node that connects each pair of the variable nodes $r_{nk}$ and $\mathbf x_{jk}$ is denoted as $f_{nk}$. Let $\Delta_{r_{nk}\leftarrow f_{nk}}(r_{nk})$ be the logarithmic form of the message from every factor node $f_{nk}$ to the corresponding variable node $r_{nk}$, while the message from $r_{nk}$ to $f_{nk}$ is denoted as $\Delta_{r_{nk}\rightarrow f_{nk}}(r_{nk})$. This notation is also used for other messages between the factor and variable nodes. In addition, we define $\Delta_{h_{km}}(h_{km})$ and $\Delta_{\mathbf x_{jk}}(\mathbf x_{jk})$ as the marginal posterior distribution of $h_{km}$ and $\mathbf x_{jk}$, respectively. Seeing Fig. \ref{sys}, the two parts
of the decoding algorithm are specified. In current channel equalization part, we consider $\exp\big(\Delta_{r_{nk}\leftarrow f_{nk}}(r_{nk})\big)$ is known, and the decoder aims to jointly update the beliefs of $\mathbf R$ and $\mathbf H$, with the prior distributions $\tilde{p}(h_{km})$ and $\exp\big(\Delta_{r_{nk}\leftarrow f_{nk}}(r_{nk})\big)$ over each $h_{km}$ and $r_{nk}$. On the other hand, for the decoding part in Sec. \ref{de}, the decoder turns to update the beliefs of $\mathbf X$, based on the messages $\Delta_{r_{nk}\rightarrow f_{nk}}(r_{nk})$. The messages $\Delta_{r_{nk}\leftarrow f_{nk}}(r_{nk})$ and $\Delta_{r_{nk}\rightarrow f_{nk}}(r_{nk})$, which alternates between the two parts of the algorithm, are formulated in Sec. \ref{cl}.

Such a problem is a bilinear inference problem \cite{big-amp}, and we have
\begin{align}
&\Delta_{g_{nm}\to r_{nk}}(r_{nk})\nonumber\\
 = &\log\int\prod\limits_{q\neq k}{\rm d}r_{nq}\prod\limits_{k,m}{\rm d}h_{km}p(y_{nm}|\sum\limits_{k}r_{nk}h_{km})\nonumber\\
\times&\prod\limits_{k}\exp\big(\Delta_{g_{nm}\leftarrow h_{km}}(h_{km})\big)\prod\limits_{q\neq k}\exp\big(\Delta_{g_{nm}\leftarrow r_{nq}}(r_{nq})\big),\\
&\Delta_{g_{nm}\leftarrow r_{nk}}(r_{nk}) = \sum\limits_{r\neq m}\Delta_{g_{nr}\to r_{nk}}(r_{nk})+\Delta_{r_{nk}\leftarrow f_{nk}}(r_{nk}),\label{rtg1}\\
&\Delta_{g_{nm}\to h_{km}}(h_{km})\nonumber\\
 = &\log\int\prod\limits_{q\neq k,m}{\rm d}h_{qm}\prod\limits_{n,k}{\rm d}r_{nk}p(y_{nm}|\sum\limits_{k}r_{nk}h_{km})\nonumber\\
\times&\prod\limits_{k}\exp\big(\Delta_{g_{nm}\leftarrow r_{nk}}(r_{nk})\big)\prod\limits_{q\neq k}\exp\big(\Delta_{g_{nm}\leftarrow h_{km}}(h_{qm})\big),\\
&\Delta_{g_{nm}\leftarrow h_{km}}(h_{km}) = \log \tilde{p}(h_{km})+\sum\limits_{v\neq n}\Delta_{g_{vm}\to h_{km}}(h_{km}).\label{htg1}
\end{align}
The marginal posterior distribution of channel coefficients can be then calculated in each iteration by
\begin{align}
\Delta_{h_{km}}(h_{km}) = &\log p(h_{km})+\sum\limits_{v\neq n}\Delta_{g_{vm}\to h_{km}}(h_{km}).
\end{align}
To handle this issue, we adopt the AMP-like approximation, which is a variant of BP that provides more efficient computations and maintains nearly Bayes-optimal performance in the large system. The core idea of AMP is to adopt the central limit theorem (CLT) to approximate the messages as Gaussian distributions, and reduce the high-order terms of messages to enhance efficiency. Since we here consider a joint update of a bilinear problem, the BiG-AMP algorithm provides an efficient solution. The key steps to integrate the BiG-AMP algorithm to handle our problem are as follow.

\begin{algorithm*}
\caption{TLMP}
\label{alg1}
\begin{algorithmic}[1]
\State $t = 1$
\State Some initializations for all the $\hat{r}_{nk}(t-1)$, $V_{nk}^r(t-1)$, $\hat{h}_{km}(t-1)$, $V_{km}^h(t-1)$, $\hat{\mathbf{x}}_{jk}(t-1)$, $\mathbf C^x_{jk}(t-1)$, $\hat{s}_{nm}(t-1)$, $\hat{t}_{nk}(t-1)$, $\hat{q}_{nk}(t-1)$, $V^q_{nk}(t-1)$, $\langle\omega_{km,g}(t-1)\rangle$,
$\bar{\gamma}^{-1}_{k,m}(t-1)$.

\Repeat

\{BiG-AMP updates
\State $\forall n,m: \bar{p}_{nm}(t) = \sum\nolimits_{k}\hat{r}_{nk}(t-1)\hat{h}_{km}(t-1)$
\State $\forall n,m: \bar{V}_{nm}^p(t)= \sum\nolimits_{k}V_{km}^h(t-1)r^2_{nk}(t-1)+V_{nk}^r(t-1)\hat{h}^2_{km}(t-1)$
\State $\forall n,m: \hat{p}_{nm}(t) =  \bar{p}_{nm}(t)-\hat{s}_{nm}(t-1)\bar{V}_{nm}^p(t)$
\State $\forall n,m: V_{nm}^p(t) = \bar{V}_{nm}^p(t)+\sum\nolimits_{k}V_{km}^h(t-1)V_{nk}^r(t-1)$
\State $\forall n,m: \hat{z}_{nm}(t) = (y_{nm}V_{nm}^p(t)+\hat{p}_{nm}(t)\sigma^2)/(V_{nm}^p(t)+\sigma^2)$
\State $\forall n,m: V_{nm}^z(t) = (V_{nm}^p(t)\sigma^2)/(V_{nm}^p(t)+\sigma^2)$
\State $\forall n,m: V_{nm}^s(t) = (1-V_{nm}^z(t)/V_{nm}^p(t))/V_{nm}^p(t)$
\State $\forall n,m: \hat{s}_{nm}(t) = (\hat{z}_{nm}(t)-\hat{p}_{nm}(t))/V_{nm}^p(t)$
\State $\forall n,k: V_{nk}^u(t) = \big(\sum\nolimits_{m}V_{nm}^s(t)\hat{h}^2_{km}(t-1)\big)^{-1}$
\State $\forall n,k: \hat{u}_{nk}(t) = \hat{r}_{nk}(t-1)(1-V_{nk}^u(t)\sum\nolimits_{m}V_{km}^h(t-1)V_{nm}^s(t))+V_{nk}^u(t)\sum\nolimits_{m}\hat s_{nm}(t)\hat h_{km}(t-1)$
\State $\forall k,m: V_{km}^w(t) = \big(\sum\nolimits_{n}V_{nm}^s(t)\hat{r}^2_{nk}(t-1)\big)^{-1}$
\State $\forall k,m: \hat{w}_{km}(t) = \hat{h}_{km}(t-1)(1-V_{km}^w(t)\sum\nolimits_{n}V_{nk}^r(t-1)V_{nm}^s(t))+V_{km}^w(t)\sum\nolimits_{n}\hat s_{nm}(t)\hat r_{nk}(t-1)$
\State $\forall n,k:\hat r_{nk}(t) = \langle r_{nk}(t);\hat{u}_{nk}(t), V_{nk}^u(t) \rangle$
\State $\forall n,k:V_{nk}^r(t) = {\rm Var}({r}_{nk}(t);\hat{u}_{nk}(t), V_{nk}^u(t))$
\State $\forall k,m: \hat h_{km}(t) =
\langle{h}_{km}(t);\hat{w}_{km}(t), V_{km}^w(t)\rangle$
\State $\forall k,m: V_{km}^h(t) = {\rm Var}({h}_{km}(t);\hat{w}_{km}(t), V_{km}^w(t))$\}

\{Ve-GAMP updates
\State $\forall n,k: {V}_{nk}^q(t)  = \sum\nolimits_{j}\mathbf{a}_{nj}^T\mathbf C_{jk}^{x}(t-1)\mathbf{a}_{nj}$
\State $\forall n,k: \hat{q}_{nk}(t) = \sum\nolimits_{j}\mathbf a^T_{nj}\hat{\mathbf x}_{jk}(t-1)-\hat{t}_{nk}(t-1){V}_{nk}^q(t)$
\State $\forall n,k: \tilde{r}_{nk}(t) = \langle{r}_{nk}(t);\hat{q}_{nk}(t),V^q_{nk}(t)\rangle$
\State $\forall n,k: \tilde{V}^r_{nk}(t) = {\rm Var}({r}_{nk}(t);\hat{q}_{nk}(t),V^q_{nk}(t))$
\State $\forall n,k: \hat{t}_{nk}(t) =  (\tilde{r}_{nk}(t)-\hat{q}_{nk}(t))/V^q_{nk}(t)$
\State $\forall n,k: V_{nk}^t(t) = (1-\tilde{V}^r_{nk}(t)/V^q_{nk}(t))/V^q_{nk}(t)$
\State $\forall j,k: \mathbf C_{jk}^{d}(t)= \big(\sum\nolimits_{n}V_{nk}^t(t)\mathbf{a}_{nj}\mathbf{a}_{nj}^T\big)^{-1}$
\State $\forall j,k: \hat{\mathbf d}_{jk}(t) = \mathbf C_{jk}^{d}(t)\sum\nolimits_{n}\hat{t}_{nk}(t)\mathbf{a}_{nj}+\hat{\mathbf{x}}_{jk}(t-1)$
\State $\forall j,k: \hat{\mathbf x}_{jk}(t) = \langle\mathbf{x}_{jk}(t);\hat{\mathbf d}_{jk}(t), \mathbf C^d_{jk}(t)\rangle$
\State $\forall j,k: \mathbf C^x_{jk}(t) = {\rm Var}(\mathbf{x}_{jk}(t);\hat{\mathbf d}_{jk}(t), \mathbf C^d_{jk}(t))$\}

\{VMP updates
\State $\forall k,m: \langle|h_{k,m}(t)|^2\rangle = \hat{h}_{km}^2(t)+V_{km}^h(t)$
\State $\forall g: \tau_g(t) = \sum\nolimits_{k,m}\langle \omega_{km,g}(t-1)\rangle+1, \quad\tilde{\tau}_g(t) = \sum\nolimits_{k,m}\sum\nolimits_{p =g+1}^G\langle \omega_{km,p}(t-1)\rangle+\alpha$
\State $\forall g: \langle\ln\beta_g(t)\rangle= \psi\left(\tau_g(t)\right)-\psi\left(\tau_g(t) +\tilde{\tau}_g(t)\right),\quad
\langle\ln(1-\beta_g(t))\rangle = \psi\left(\tilde{\tau}_g(t))-\psi(\tau_g(t) +\tilde{\tau}_g(t)\right)$
\If{the reduction condition for component $g$ is satisfied}
\State Remove the component $g$, $G = G-1$
\EndIf
\State $\forall g,m: \tilde{a}_{g}(t) = \sum\nolimits_{k,m}\frac{1}{2}\langle \omega_{km,g}(t-1)\rangle+a, \quad\tilde{b}_{g}(t) = \sum\nolimits_{k,m}\frac{1}{2}\langle \omega_{km,g}(t-1)\rangle\langle|h_{k,m}(t)|^2\rangle+b$
\State $\forall g: \langle\gamma_{g}(t)\rangle = \tilde{a}_{g}(t)\big(\tilde{b}_{g}(t)\big)^{-1},\quad \langle\ln\gamma_{g}(t)\rangle = \psi(\tilde{a}_{g}(t))-\ln\tilde{b}_{g}(t)$
\State $\forall k,m,g: \langle\omega_{km,g}(t)\rangle = \zeta_{km,g}(t)(\sum\nolimits_{g}\zeta_{km,g}(t))^{-1}$ with $\zeta_{km,g}(t)$ defined in (\ref{lnzeta})
\State $\forall k,g: \bar{\gamma}_{km}(t)= \sum\nolimits_{g}\langle \omega_{km,g}(t)\rangle\langle\gamma_{g}(t)\rangle$\}
\State $t = t+1$
\Until{$\sum\nolimits_{j,k}||\hat{\mathbf x}_{jk}(t)-\hat{\mathbf x}_{jk}(t-1)||^2>\epsilon\sum\nolimits_{j,k}||\hat{\mathbf x}_{jk}(t-1)||^2$ or $t>T_{\rm max}$}
\end{algorithmic}
\end{algorithm*}
Before proceeding, we define another marginal posterior distribution for each of the codeword $r_{nk}$ as
$
\Delta_{r_{nk}}(r_{nk}) = \sum\limits_{m}\Delta_{g_{nm}\to r_{nk}}(r_{nk})+\Delta_{r_{nk}\leftarrow f_{nk}}(r_{nk})
$.
The mean and variance associated with $\Delta_{h_{km}}(h_{km})$ and $\Delta_{r_{nk}}(r_{nk})$ are denoted as $\hat{h}_{km}$, $V_{km}^h$ and $\hat{r}_{nk}$, $V_{nk}^r$, respectively. The code idea of the BiG-AMP is to jointly update these statistics $\hat{h}_{km}$, $V_{km}^h$ and $\hat{r}_{nk}$, $V_{nk}^r$ with some initializations. The update equations are shown in Algorithm \ref{alg1}, and the detailed derivations of the BiG-AMP are provided in the Appendix \ref{1}. The mean and variance $\langle{r}_{nk};\hat{u}_{nk}, V_{nk}^u\rangle$, and ${\rm Var}({r}_{nk};\hat{u}_{nk}, V_{nk}^u)$ in Algorithm \ref{alg1} are taken over the PDF
\begin{align}
p({r}_{nk};&\hat{u}_{nk}, V_{nk}^u) \nonumber\\
&= \frac{1}{C}\exp\big(\Delta_{r_{nk}\leftarrow f_{nk}}(r_{nk})\big)\mathcal{N}(r_{nk};\hat{u}_{nk}, V_{nk}^u).\label{r_pos}
\end{align}
The term $\exp\big(\Delta_{r_{nk}\leftarrow f_{nk}}(r_{nk})\big)$ will be specified in the Sec. III-C.
Similarly, $\langle{h}_{km};\hat{w}_{km}, V_{km}^w\rangle$, and ${\rm Var}({h}_{km};\hat{w}_{km}, V_{km}^w)$ in Algorithm \ref{alg1} are taken over the PDF
\begin{align}
p({h}_{km};\hat{w}_{km}, V_{km}^w) = \frac{1}{C}\tilde{p}(h_{km})\mathcal{N}(h_{km};\hat{w}_{km}, V_{km}^w).
\end{align}
Based on (\ref{pri_h_tran}), we therefore have
\begin{align}
\hat{h}_{km}= \frac{\hat{w}_{km}\bar{\gamma}_{km}^{-1}}{\bar{\gamma}_{km}^{-1}+V_{km}^w},\quad V_{km}^h = \frac{\bar{\gamma}_{km}^{-1}V_{km}^w}{\bar{\gamma}_{km}^{-1}+V_{km}^w}.
\end{align}
We note that the derivations use a general Gaussian assumption of the linear mixing variables, defined as $z_{nm} = r_{nk}h_{km}+\sum\nolimits_{q\neq k}r_{nq}h_{qm}$. The Gaussianity of such variables are clarified as follow. The length of each variable $\mathbf x_{jk}$ is exponentially related to the length $L$ of the sub-sequences, so that even a moderate $L$ will cause it far larger than one. This results in that the probability that two of the devices transmit a same sub-sequence in a block will be much less than one, i.e. $1/2^L$. We therefore can safely treat the elements in the codeword matrix $\mathbf R$ as mutually independent random variables. Combining the factorized form of the prior distribution over $\mathbf H$, the CLT can be adopted, so that the linear mixing variables can be treated as Gaussian random variables.

As aforementioned, we have updated the marginal posterior distributions of the variables $\mathbf R$, $\mathbf H$, $\boldsymbol\Omega$, $\boldsymbol\gamma$ and $\boldsymbol\beta$, based on the message $\Delta_{r_{nk}\leftarrow f_{nk}}(r_{nk})$ in the channel equalization part of our algorithm. After that, we can return the messages $\Delta_{r_{nk}\rightarrow f_{nk}}(r_{nk})$. In the next decoding part, our task turns to
update the marginal posterior distribution of $\mathbf X$, based on the messages $\Delta_{r_{nk}\rightarrow f_{nk}}(r_{nk})$.
\subsection{Part II: Decoding}\label{de}
Recalling Fig. \ref{sys}, the factor graph corresponding to the decoding part can be regarded as a linear mixing problem, where $\exp\big(\Delta_{r_{nk}\rightarrow f_{nk}}(r_{nk})\big)$ is the equivalent output distribution over each linear mixing variable $r_{nk}$, defined by $r_{nk} = \sum\limits_{j}\mathbf a_{nj}^T\mathbf x_{jk}$. We here extend the traditional GAMP algorithm \cite{GAMP} into the vector-valued case, called vector-valued GAMP (Ve-GAMP), where each the variable $r_{nk}$ is linearly mixed by a series of vector-valued random variables $(\mathbf x_{1k}|\dots|\mathbf x_{Jk})$. Based on Fig. \ref{sys}, we have
\begin{align}
&\Delta_{f_{nk}\to \mathbf x_{jk}}(\mathbf x_{jk})\nonumber\\
 = &\log\int\prod\nolimits_{p\neq j,k}{\rm d}\mathbf x_{jk}\prod\nolimits_{n,k}{\rm d}r_{nk}\delta(r_{nk} = \sum\nolimits_{j}\mathbf a_{nj}^T\mathbf x_{jk})\nonumber\\
\times&\exp\big(\Delta_{r_{nk}\rightarrow f_{nk}}(r_{nk})\big)\prod\nolimits_{p\neq j}\exp\big(\Delta_{f_{nk}\leftarrow \mathbf x_{jk}}(\mathbf x_{jk})\big),\label{ftxo}
\end{align}
\begin{align}
\Delta_{f_{nk}\leftarrow \mathbf x_{jk}}(\mathbf x_{jk}) = \log p(\mathbf x_{jk})+ \sum\nolimits_{v\neq n}\Delta_{f_{nk}\to \mathbf x_{jk}}(\mathbf x_{jk}).\label{xtf1}
\end{align}
The marginal posterior distribution of each the variable $\mathbf x_{jk}$ is then calculated by
\begin{align}
\Delta_{\mathbf x_{jk}}(\mathbf x_{jk}) = &\log p(\mathbf x_{jk})+ \sum\nolimits_{v\neq n}\Delta_{f_{nk}\to \mathbf x_{jk}}(\mathbf x_{jk}), \label{pos_x}
\end{align}
with the corresponding mean and covariance matrix defined as $\hat{\mathbf x}_{jk}$ and $\mathbf C^x_{jk}$, respectively.
Ve-GAMP is then to update the statistics
$\hat{\mathbf x}_{jk}$ and $\mathbf C^x_{jk}$ in each iteration. The detailed derivations of the Ve-GAMP are provided in Appendix \ref{2}. The obtained update equations are shown in Algorithm \ref{alg1}.
The mean and covariance matrix $\langle\mathbf{x}_{jk};\hat{\mathbf d}_{jk}, \mathbf C^d_{jk}\rangle$, and ${\rm Var}(\mathbf{x}_{jk};\hat{\mathbf d}_{jk}, \mathbf C^d_{jk})$ are taken over the PDF
\begin{align}
p(\mathbf{x}_{jk};\hat{\mathbf d}_{jk}, \mathbf C^d_{jk}) = \frac{1}{C}p(\mathbf x_{jk})\mathcal{N}(\mathbf x_{jk}; \hat{\mathbf d}_{jk}, \mathbf C_{jk}^{d}).\label{pos_xv}
\end{align}
Besides, $\langle{r}_{nk};\hat{q}_{nk},V^q_{nk}\rangle$ and ${\rm Var}({r}_{nk};\hat{q}_{nk},V^q_{nk})$ are taken over the PDF
\begin{align}
p({r}_{nk};&\hat{q}_{nk},V^q_{nk})\nonumber\\
= &\frac{1}{C}\exp\big(\Delta_{r_{nk}\rightarrow f_{nk}}(r_{nk})\big)\mathcal{N}(r_{nk};\hat{q}_{nk},V^q_{nk}).\label{posrq}
\end{align}
Note that the term $\exp\big(\Delta_{r_{nk}\rightarrow f_{nk}}(r_{nk})\big)$ will be specified in Sec. III-C.

\begin{figure}[!t]
\centering
\includegraphics[width=3in]{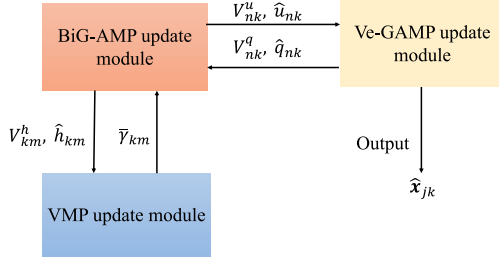}
\caption{A flow diagram of Algorithms 1.}
\label{alomodu}
\end{figure}
\subsection{Closing the Loop}\label{cl}
In the previous sections, we almost deduce the whole decoding algorithm TLMP. The ultimate step to complete TLMP is to specify the messages $\Delta_{r_{nk}\leftarrow f_{nk}}(r_{nk})$ and $\Delta_{r_{nk}\rightarrow f_{nk}}(r_{nk})$, which alternates between the two parts of the algorithm. The message $\Delta_{r_{nk}\leftarrow f_{nk}}(r_{nk})$ can be considered as the prior distribution over each $r_{nk}$. As aforementioned, we adopt the Gaussian approximation to each of the variable $r_{nk}$, whose mean and variance is determined by the messages $\Delta_{f_{nk}\leftarrow \mathbf{x}_{jk}}(r_{nk})$. Inferring from the derivations in Appendix \ref{2}, we can straightforward obtain
\begin{equation}\Delta_{r_{nk}\leftarrow f_{nk}}(r_{nk}) = \mathcal{N}(r_{nk};\hat{q}_{nk}, {V}_{nk}^q).\label{rtf}\end{equation} On the other hand, the message $\Delta_{r_{nk}\rightarrow f_{nk}}(r_{nk})$ can be obtained by performing the general BP as
$
\Delta_{r_{nk}\rightarrow f_{nk}}(r_{nk}) = \sum\limits_{m}\Delta_{g_{nm}\rightarrow r_{nk}}(r_{nk})
$.
Based on the derivations on Appendix \ref{1}, we have \begin{equation}\Delta_{r_{nk}\rightarrow f_{nk}}(r_{nk}) = \mathcal{N}(r_{nk}; \hat{u}_{nk}, V_{nk}^u).\label{ftr}\end{equation} Combining (\ref{rtf}), (\ref{ftr}) with (\ref{r_pos}) and (\ref{posrq}), the mean and variance of (\ref{r_pos}) can be obtained, given by
\begin{align}
\hat{r}_{nk} = \frac{\hat{u}_{nk}V_{nk}^q+\hat{q}_{nk}V_{nk}^u}{V_{nk}^q+V_{nk}^u},\quad V_{nk}^r = \frac{V_{nk}^qV_{nk}^u}{V_{nk}^q+V_{nk}^u}.\label{r_mv}
\end{align}
We also notice that the mean and variance of (\ref{posrq}) are updated in the same way as (\ref{r_mv}). The variable $\mathbf R$ can be regarded as the intermediate variable of the TLMP, i.e, the part I of TLMP updates $\Delta_{r_{nk}\rightarrow f_{nk}}(r_{nk})$ and transmits to the part II as the likelihood of $\mathbf R$ and the part II of TLMP updates the corresponding prior distribution $\Delta_{r_{nk}\leftarrow f_{nk}}(r_{nk})$ and transmits it to the part I. The basic steps of the TLMP can be summarized as follows. After initializations{\footnote {For initializing the mean of $\mathbf H$, we produce a piece of additional bits encoded by CS codes and execute the CS decoding algorithm in \cite{Shyianov2021} with the collision protocol \cite{2022Li} to derive a rough estimated channel with non-codewords collisions.}}, the decoder updates the means $\hat{u}_{nk}$, $\hat{h}_{km}$ and variances $V_{nk}^u$, $V_{km}^h$ of each $r_{nk}$ and $h_{km}$ via the BiG-AMP module. Then, the decoder propagates $\hat{u}_{nk}$ and $V_{nk}^u$ into Ve-GAMP module. In the meanwhile, the decoder propagates $\hat{h}_{km}$ and $V_{km}^h$ into VMP module.
After the Ve-GAMP updates, the updated mean $\hat{q}_{nk}$ and variance $V_{nk}^q$ of each $r_{nk}$ are fed back to the BiG-AMP module as the prior parameters for the subsequent updates. After the VMP updates, the updated variance $\bar{\gamma}_{km}^{-1}$ of each $h_{km}$ are also fed back to the BiG-AMP module. For the conciseness, we provide a flow diagram of Algorithms 1 with individual boxes for the components in Fig. \ref{alomodu}.

\begin{algorithm*}
\caption{TLMP(modified)}
\label{alg2}
\begin{algorithmic}[1]
\State $t = 1$
\State Initializations that will be specified in Sec. V
\Repeat

\{BiG-AMP updates
\State $\bar{\mathbf P}(t) = \hat{\mathbf R}(t-1)\hat{\mathbf H}(t-1)$
\State $\bar{V}^p(t) = \theta_1(1/MV^r(t-1)||\hat{\mathbf H}(t-1)||^2_F+1/NV^h(t-1)||\hat{\mathbf R}(t-1)||^2_F)+(1-\theta_1)\bar{V}^p(t-1)$
\State $V^p(t) = \theta_1(\bar{V}^p(t)+KV^r(t-1)V^h(t-1))+(1-\theta_1)V^p(t-1)$
\State $\hat{\mathbf P}(t) = \bar{\mathbf P}(t)-\bar{V}^p(t)\hat{\mathbf S}(t-1)$
\State $\hat{\mathbf S}(t) = \theta_1(\mathbf Y-\hat{\mathbf P}(t))/(V^p(t)+\sigma^2)+(1-\theta_1)\hat{\mathbf S}(t-1)$
\State $V^s(t) = \theta_1/(V^p(t)+\sigma^2)+(1-\theta_1)V^s(t-1)$
\State $V^u(t) = K/(V^s(t)||\bar{\mathbf H}(t-1)||^2_F)$
\State $\hat{\mathbf U}(t) = (1-MV^u(t)V^s(t)V^h(t-1))\bar{\mathbf R}(t-1)+V^u(t)\hat{\mathbf S}(t)\bar{\mathbf H}^T(t-1)$
\State $V^w(t) = K/(V^s(t)||\bar{\mathbf R}(t-1)||^2_F)$
\State $\hat{\mathbf W}(t) = (1-NV^w(t)V^s(t)V^r(t-1))\bar{\mathbf H}(t-1)+V^w(t)\bar{\mathbf R}^T(t-1)\hat{\mathbf S}(t)$
\State $\forall n,k:\hat r_{nk}(t) = \langle{r}_{nk};\hat{u}_{nk}(t), V^u(t)\rangle$
\State $\bar{\mathbf R}(t) = \theta_1\hat{\mathbf R}(t)+(1-\theta_1)\bar{\mathbf R}(t-1)$
\State $\forall n,k:V^r(t) = 1/(NK)\sum\nolimits_{n,k}{\rm Var}({r}_{nk};\hat{u}_{nk}(t), V^u(t))$
\State $\forall k,m: \hat h_{km}(t) = \langle{h}_{km};\hat{w}_{km}(t), V^w(t)\rangle$
\State $\bar{\mathbf H}(t) = \theta_1\hat{\mathbf H}(t)+(1-\theta_1)\bar{\mathbf H}(t-1)$
\State $\forall k,m: V^h(t) = 1/(KM)\sum\nolimits_{k,m}{\rm Var}({h}_{km};\hat{w}_{km}(t), V^w(t))$\}

\{Ve-GAMP updates
\State ${V}^q(t)  = \theta_2(1/N||\mathbf A||^2_FV^x(t-1)+(1-\theta_2){V}^q(t-1)$
\State $\hat{\mathbf Q}(t) = \mathbf A\hat{\mathbf X}(t-1)-{V}^q(t)\hat{\mathbf T}(t-1)$
\State $\hat{\mathbf R}(t) = ({V}^q(t)\hat{\mathbf U}(t)+{V}^u(t)\hat{\mathbf Q}(t))/({V}^q(t)+{V}^u(t))$
\State $V^r(t) = {V}^q(t){V}^u(t)/({V}^q(t)+{V}^u(t))$
\State $\hat{\mathbf T}(t) = \theta_2(\hat{\mathbf U}(t)-\hat{\mathbf Q}(t))/({V}^q(t)+{V}^u(t))+(1-\theta_2)\hat{\mathbf T}(t-1)$
\State $V^t(t) = \theta_2/({V}^q(t)+{V}^u(t))+(1-\theta_2)V^t(t-1)$
\State $V^d(t) = J2^L/(V^t(t)||\mathbf A||^2_F)$
\State $\hat{\mathbf D}(t) = \bar{\mathbf X}(t-1)+V^d(t)\mathbf A^T\hat{\mathbf T}(t)$
\State $\forall j,k: \hat{\mathbf x}_{jk}(t) = \langle\mathbf{x}_{jk};\hat{\mathbf D}(t), V^d(t)\rangle$
\State $\forall j,k: \bar{\mathbf x}_{jk}(t) = \theta_2\hat{\mathbf x}_{jk}(t)+(1-\theta_2)\bar{\mathbf x}_{jk}(t-1)$
\State $V^x(t) = 1/(2^LJK)\sum\nolimits_{j,k,l}{\rm Var}(\mathbf{x}_{jk,l};\hat{d}_{jk,l}(t), V^d(t))$\}

\{VMP updates in Algorithm \ref{alg1}\}
\State $t = t+1$
\Until{$\sum\nolimits_{j,k}||\hat{\mathbf x}_{jk}(t)-\hat{\mathbf x}_{jk}(t-1)||^2>\epsilon\sum\nolimits_{j,k}||\hat{\mathbf x}_{jk}(t-1)||^2$ or $t>T_{\rm max}$}
\end{algorithmic}
\end{algorithm*}
\section{The Implementation Details of the Decoding Algorithm}
We in this section provide some implementation details of the proposed TLMP so that it can be efficiently used in practical applications of interest. We first describe some simplifications of TLMP to further reduce the computational complexity based on the uses of scalar variances.
To avoid the diverge, we then propose to incorporate the damping into TLMP, which has been successfully used in the classical AMP-based methods to provide convergence guarantees \cite{Schniter2015, big-amp}. In addition, we introduce a cost function to evaluate and monitor the performance of TLMP in each iteration.

\subsection{Decoding Algorithm Simplifications}
TLMP stores and processes a number of element-wise variance terms with the corresponding values varying across elements. To further reduce the memory and complexity of the algorithm, the variance terms are replaced with scalars, which forces all the variance components to be the same. We note that the covariance matrix $\mathbf C_{jk}^{d}$ and $\mathbf C_{jk}^{x}$ are approximated as diagonal with the same diagonal elements $V^d$ and $V^x$, since the independently drawn elements in $\mathbf A$ make the non diagonal terms of the covariance matrices of a smaller order than the diagonal ones.
Using such simplifications, we can approximate the mean $\hat{\mathbf{x}}_{jk} = (\hat{{x}}_{jk,1}|\dots|\hat{{x}}_{jk,2^L})$ and variance terms $V^x$ of (\ref{pos_xv}) as
\begin{align}
\hat{{x}}_{jk,l} &= \exp\left(-\frac{1-2\hat{{d}}_{jk,l}}{2V_d}\right)\left(\sum\nolimits_{l}\exp\left(-\frac{1-2\hat{{d}}_{jk,l}}{2V_d}\right)\right)^{-1},\nonumber\\
V^x &= \frac{1}{JK2^L}\sum\nolimits_{j,k,l}\hat{{x}}_{jk,l}(1-\hat{{x}}_{jk,l})\nonumber,
\end{align}
where $\hat{{d}}_{jk,l}$ is the $l$th element of $\hat{\mathbf d}_{jk}$. The modified equations are listed in Algorithm \ref{alg2}. We note that we use the matrix forms of the involved variables for the ease of expressions. After that, we execute a symbol-by-symbol judgment for each $\mathbf x_{jk}$, and the maximum element in the estimation of $\mathbf x_{jk}$ is regarded as the non-zero location of $\mathbf x_{jk}$.

The complexity of the modified TLMP can be analyzed as follow. In each iteration, the complexity of the VMP updates is $\mathcal O(KMG)$, the complexity of the BiG-AMP updates is $\mathcal O(NKM)$ and the complexity of the Ve-AMP updates is $\mathcal O(NKJ2^L)$. Since the value of $G$ always much smaller than other system parameters, the overall complexity of our algorithm is dominated by the AMP-like algorithms, which is $\mathcal O(NK(J2^L+M))$. Note that the complexity increases linearly with the increases of each of the parameters, which is similar with the standard AMP algorithm.

\subsection{Decoding Algorithm Damping}
To avoid the misconvergence of our algorithm caused by the finite dimensions, it is helpful to damp the iterations, which reduce the speed of parameters change in the algorithm. This is achieved by using damping factor $\theta_1$ for the channel equalization part and $\theta_2$ for the decoding part. Based on the damping factors, some newly provided equations are then used in place of the original update steps, as can be seen in Algorithm \ref{alg2}. We note that when the factors $\theta_1 = \theta_2 = 1$, the damping has no effect, whereas when $\theta_1 = \theta_2 = 0$, all quantities become frozen. To adaptively chose the  appropriate damping factors, we refer to \cite{big-amp}, where a cost criterion is established and the factors can be adaptively adjusted by monitoring the criterion.

\subsection{Performance Evaluation}
We in this section establish a cost function, used to measure the similarity between the estimated posterior distribution by the decoding algorithm and the exact posterior distribution $p(\boldsymbol\Theta|\mathbf Y)$ inferred from (\ref{jot_dis}). The cost function can be used in each iteration to evaluate the algorithm performance as well as to adjust the damping factors. We denote the estimated posterior distribution as
$
\tilde{q(\boldsymbol\Theta)} = \tilde{q}(\mathbf H, \mathbf R, \mathbf X)q(\boldsymbol\Omega)q(\boldsymbol\beta)q(\boldsymbol\gamma)
$,
where the $q(\boldsymbol\Omega)$, $q(\boldsymbol\beta)$ and $q(\boldsymbol\gamma)$ have been derived in (\ref{qome}), (\ref{qbet}) and (\ref{qgam}) based on VMP, respectively. The distribution $\tilde{q}(\mathbf H, \mathbf R, \mathbf X)$ is an AMP estimation version of (\ref{tran_mod}), with the marginal distributions written as $
\tilde{q}(\mathbf H)\propto \prod_{k,m} \tilde{p}(h_{km})\mathcal{N}(h_{km};\hat{w}_{km}, V_{km}^w)$,
$\tilde{q}(\mathbf R) \propto \prod_{n,k}\mathcal{N}(r_{nk}; \hat{u}_{nk}, V_{nk}^u)\mathcal{N}(r_{nk};\hat{q}_{nk},V^q_{nk})$, $
\tilde{q}(\mathbf X)\propto p(\mathbf x_{jk})\mathcal{N}(\mathbf x_{jk}; \hat{\mathbf d}_{jkn}, \mathbf C_{jkn}^{d})$.\footnote{ After the scalar value approximations, the variance term can be considered as scales, and the covariance matrix $\mathbf C_{jkn}^{d}$ is a diagonal matrix with same diagonal elements $V^d$.} Similarly with Sec. \ref{kl}, we use the KL divergence to measure the similarity between $\tilde{q(\boldsymbol\Theta)}$ and the exact posterior distribution, which is given by
\begin{align}
&\mathcal{D}_{{\rm KL}}[\tilde{q}(\boldsymbol\Theta)||p(\boldsymbol\Theta|\mathbf Y)]\nonumber\\
 =  &\langle\ln\tilde{q}(\mathbf H, \mathbf R, \mathbf X)\rangle_{\tilde{q}(\mathbf H, \mathbf R, \mathbf X)}+\langle\ln q(\boldsymbol\Omega)q(\boldsymbol\beta)q(\boldsymbol\gamma)\rangle_{q(\boldsymbol\Omega)q(\boldsymbol\beta)q(\boldsymbol\gamma)}\nonumber\\
 &\quad\quad\quad\quad-\langle\ln p(\boldsymbol\Theta, \mathbf Y)\rangle_{\tilde{q}(\boldsymbol\Theta)}+{\rm const}\label{klqp1}\\
 = &\mathcal{D}_{{\rm KL}}[\tilde{q}(\mathbf H, \mathbf R, \mathbf X)||{q}(\mathbf H, \mathbf R, \mathbf X)]+{\rm const}\nonumber\\
 &\quad\quad\quad\quad+\mathcal{D}_{{\rm KL}}[q(\boldsymbol\Omega)q(\boldsymbol\beta)q(\boldsymbol\gamma)||
 p(\boldsymbol\Omega|\boldsymbol\beta)p(\boldsymbol\gamma)p(\boldsymbol\beta)]\label{klqp2}\\
 = &\mathcal{D}_{{\rm KL}}[\tilde{q}(\mathbf H, \mathbf R, \mathbf X)||p(\mathbf R|\mathbf X)p(\mathbf X)\tilde{p}(\mathbf H)]\nonumber\\
 &\quad\quad\quad\quad+\mathcal{D}_{{\rm KL}}[q(\boldsymbol\Omega)q(\boldsymbol\beta)q(\boldsymbol\gamma)||
 p(\boldsymbol\Omega|\boldsymbol\beta)p(\boldsymbol\gamma)p(\boldsymbol\beta)]\nonumber\\
 &\quad\quad\quad\quad-\langle\ln p(\mathbf Y|\mathbf H, \mathbf R)\rangle_{\tilde{q}(\mathbf H)\tilde{q}(\mathbf R)}+{\rm const}\label{klqp3}\\
  = &\mathcal{D}_{{\rm KL}}[\tilde{q}(\mathbf H)||\tilde{p}(\mathbf H)]+\mathcal{D}_{{\rm KL}}[\tilde{q}(\mathbf X)||p(\mathbf X)]\nonumber\\
   &\quad\quad\quad\quad+\mathcal{D}_{{\rm KL}}[q(\boldsymbol\Omega)q(\boldsymbol\beta)q(\boldsymbol\gamma)||
 p(\boldsymbol\Omega|\boldsymbol\beta)p(\boldsymbol\gamma)p(\boldsymbol\beta)]\nonumber\\
 &\quad\quad\quad\quad-\langle\ln p(\mathbf Y|\mathbf H, \mathbf R)\rangle_{\tilde{q}(\mathbf H)\tilde{q}(\mathbf R)}+{\rm const}\label{klqp4}.
\end{align}
The equation (\ref{klqp1}) is derived from the definition of the KL divergence, where the items that are independent of $\mathbf H$, $\mathbf R$ and $\mathbf X$ are absorbed into the constant term. The equation (\ref{klqp2}) uses the general VMP principle in (\ref{log_opt}). The equation (\ref{klqp3}) combines the probabilistic model (\ref{tran_mod}), and the equation (\ref{klqp4}) considers the constraint $\mathbf R = \mathbf A\mathbf X$. To further calculate the term $\langle\ln p(\mathbf Y|\mathbf H, \mathbf R)\rangle_{\tilde{q}(\mathbf H)\tilde{q}(\mathbf R)}$ in (\ref{klqp4}), we can approximately use an independent Gaussian matrix $\mathbf Z$ in place of $\mathbf R\mathbf H$, and the cost function can be obtained immediately.
The cost function can be used in each iteration to evaluate the algorithm performance. When the cost has not decreased sufficiently, it means current  iteration does not increase the similarity between the estimated posterior distribution and the exact one, and we can decrease the damping factors and rerun the loop.
\begin{figure}[ht]
\subfigure[With codewords collisions.]{
\begin{minipage}[b]{0.48\textwidth}
\includegraphics[width=3.2 in]{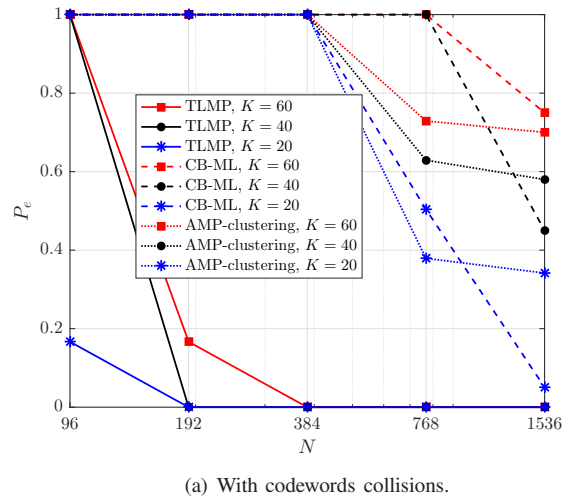}
\end{minipage}
}
\subfigure[Without codewords collisions.]{
\begin{minipage}[b]{0.48\textwidth}
\includegraphics[width=3.2 in]{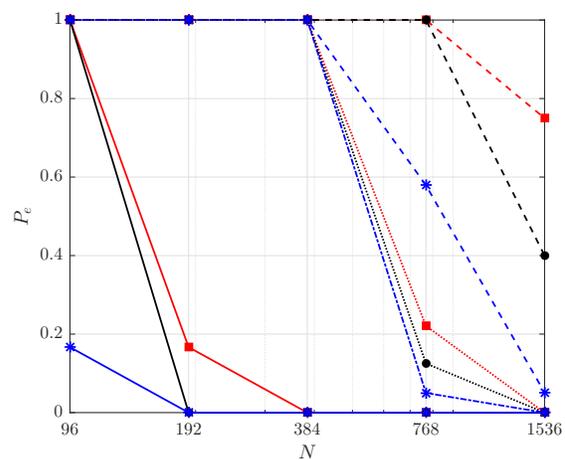}
\end{minipage}
}
\caption{Error probability $P_e$ as a function of the length of codewords $N$, and the number of users $K$, in the cases with or without codewords collisions.}
\label{com_coll}
\end{figure}

\section{Numerical Results}
In this section, we provide some numerical examples to verify our theoretical results. We simulate the mMTC system with a BS equipped with an uniform antenna array with $M$ antennas. We choose $B$ bits as payload size for each user. The SNR in dB is defined as $10\log1/\sigma^2$. The initializations of the TLMP are described as follow. we set $\hat{\mathbf R}(0) = \bar{\mathbf R}(0) = \mathbf A\hat{\mathbf X}(0)$, where $\hat{\mathbf X}(0) = \bar{\mathbf X}(0)$ is randomly generated by (\ref{pri_x}). We set $V^r(0)$ and $V^x(0)$ as $10$ times over their prior variance, $V^h(0) = 10$ and ${V}^q(0)= 1/N$. $\bar{V}^p(0)$, $V^p(0)$, $\hat{\mathbf S}(0)$, $V^s(0)$, $\hat{\mathbf Q}(0)$, $\hat{\mathbf T}(0)$, $V^t(0)$ are set as zero. Terms $\langle\omega_{km,g}(0)\rangle$ are uniformly produced  based on their prior distribution. All the terms $\bar{\gamma}^{-1}_{k,m}(0)$ are set as $10$. To initialize $\hat{\mathbf H}(0)$, we produce a piece of additional bits of length $L_0$, and use the CS encoding with length of the codewords $N_0$. The estimated channel is $\hat{\mathbf H}(0)$, and we set $\bar{\mathbf H}(0) =  \hat{\mathbf H}(0)$. We note that we will not use the collision protocol anywhere else, and the additional bits are not considered as the information bits for fair. The overall spectral efficiency of the TLMP is formulated as $BK/(N+N_0)$.

For comparisons, we first compare our TLMP algorithm with the two divide-and-conquer baselines, i.e., the AMP-clustering algorithm in \cite{Shyianov2021}, and the CB-ML algorithm in \cite{2019ale}. We here consider the classical one ring channel model in \cite{JSDM2013} for each of the user devices, which captures the spatially correlations between channel coefficients in the practice. The azimuth angle of each user device is generated randomly and the angular spread (AS) is set as $40^\circ$. The large scale fading coefficients for all the users, i.e., shadowing and path-loss, are assumed to be a constant that is absorbed into the noise variance.

We set $B = 96$ bits as payload size for each user. We set $N_0 = K$, $L_0 = 10$ for the initialization. The number of blocks for our TLMP and AMP-clustering equals to $J = B/L$. Differently, for the CB-ML, there are extra parity bits, with the number set as three times larger than the information bits, which is referred to \cite{2019ale}. For the TLMP, the number of bits in each section is set as $L = 8$. The SNR is set as $20$dB. Particularly, for the initialization of our TLMP, we use a $0$dB SNR to obtain a very rough estimated channel.
\begin{figure}[!h]
\centering
\includegraphics[width=3.2in]{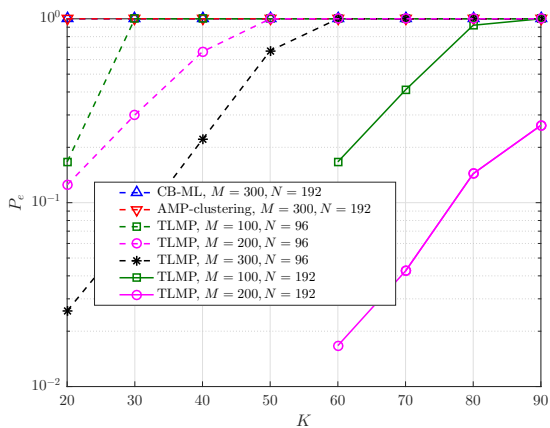}
\caption{Error probability $P_e$ as a function of the number of users $K$, the length of codewords $N$, the number of BS antennas $M$.}
\label{KM}
\end{figure}
\begin{figure}[ht]
\subfigure[$N = 80$, $B = 60$.]{
\begin{minipage}[b]{0.48\textwidth}
\includegraphics[width=3.2 in]{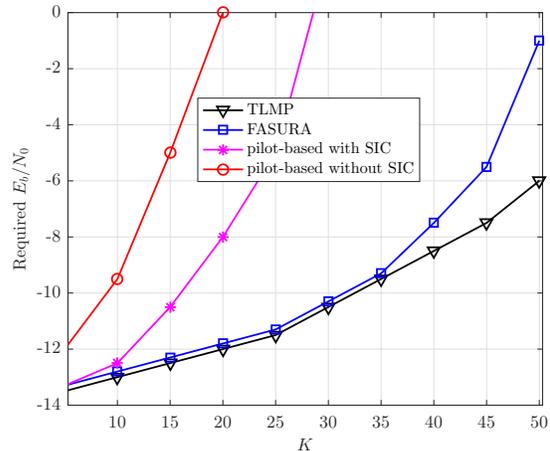}
\end{minipage}
}
\subfigure[$N = 160$, $B = 100$.]{
\begin{minipage}[b]{0.48\textwidth}
\includegraphics[width=3.2 in]{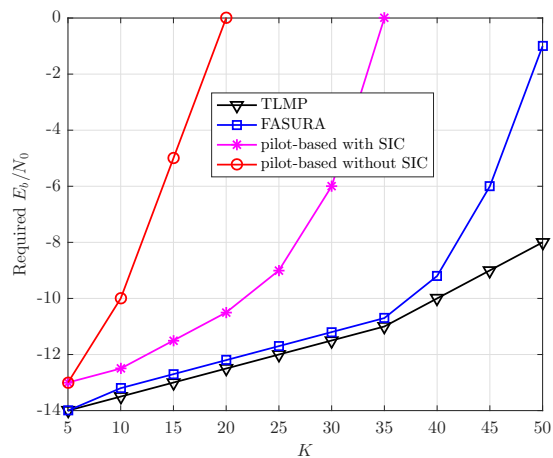}
\end{minipage}
}
\caption{$E_b/N_0$ versus $K$ curves of TLMP and the typical pilot-based methods in different settings of $N$, $B$ with $M = 50$.}
\label{ebn01}
\end{figure}
Fig. \ref{com_coll} depicts the error probabilities of our proposed TLMP and its counterparts. The number of the antenna of BS is set as $M = 100$. To emphasize the robustness of the TLMP dealing with the possible codewords collisions, we demonstrate two cases. In the first case with codewords collisions, we produce the codewords uniformly based on the prior distribution (\ref{pri_x}) and the codebook $\mathbf A$. In the second case without codewords collisions, we artificially limit the generations of the codewords to ensure that there are no codewords collisions. We can see from the Fig. \ref{com_coll} that our proposed TLMP significantly enhances the spectral efficiency compared with the CB-ML and AMP-clustering baselines to achieve negligible error probability in the case without the codewords collisions. We can also observe from Fig. \ref{com_coll} that the codewords collisions will damage the performance of AMP-clustering algorithm, while have negligible impacts on our TLMP. Compared with the CB-ML, the performance gain of our TLMP mainly comes from that we do not require the redundant parity bits to be inserted aiming to stitch all the sequences together, which can be achieved by better using the massive MIMO channels. Compared with the AMP-clustering, we clearly see the benefit of jointly decoding all the blocks rather than considering each block as an independent CS instance.

Fig. \ref{KM} depicts the error probabilities of our proposed TLMP and its counterparts as a function of the number of users $K$, the length of codewords $N$, the number of BS antennas $M$. We can see an apparent advantage of the TLMP than the CB-ML and AMP-clustering. We also find that with fixed length of codewords $N$, increasing $M$ will increase the affordable number of users $K$ under certain error probability threshold. This hinds the benefits of using the massive MIMO. In addition, we find that increasing $N$ will bring more performance compared with increasing $M$ with the same $K$. This is because the codewords length $N$ is beneficial to both the two parts of the TLMP, while the number $M$ only has an impact on the channel equalization part.

\begin{figure}[ht]
\subfigure[$N = 80$, $B = 60$.]{
\begin{minipage}[b]{0.48\textwidth}
\includegraphics[width=3.2 in]{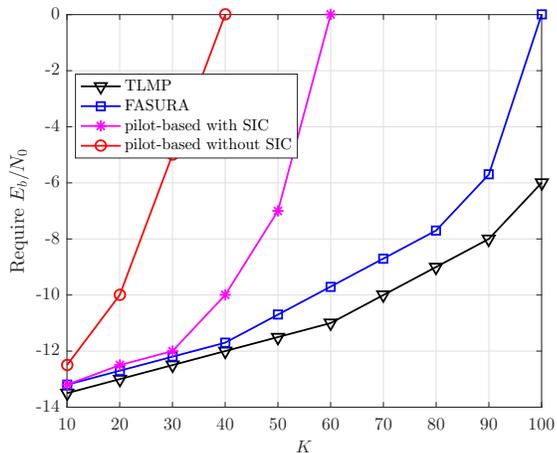}
\end{minipage}
}
\subfigure[$N = 160$, $B = 100$.]{
\begin{minipage}[b]{0.48\textwidth}
\includegraphics[width=3.2 in]{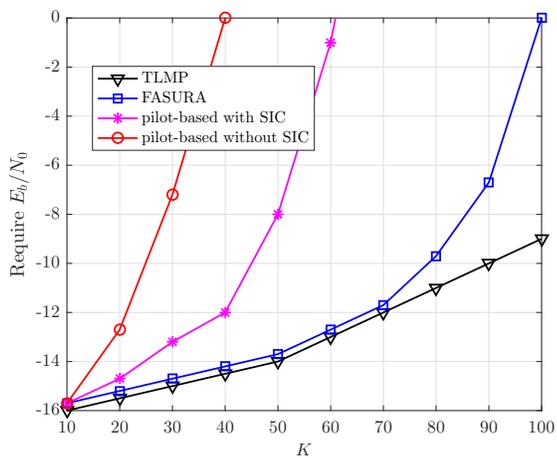}
\end{minipage}
}
\caption{$E_b/N_0$ versus $K$ curves of TLMP and the typical pilot-based methods in different settings of $N$, $B$ with $M = 100$.}
\label{ebn02}
\end{figure}

We then compare our TLMP algorithm with the typical pilot-based algorithms. We choose three typical pilot-based baselines in  \cite{PilotFengler, PilotAhmadi, FASURA}, which are referred as \emph{pilot-based method without SIC}, \emph{pilot-based method with SIC} and \emph{FASURA}, respectively. We adopt the Rayleigh channel model with zero mean and unit variance here, since the competitors are dependent on such a channel model. Figs. \ref{ebn01} and \ref{ebn02} demonstrate the $E_b/N_0$ versus $K$ curves of our TLMP and its pilot-based baselines, where the $E_b/N_0$ in dB is the energy-per-bit of the system. We set $L = 5$ for TLMP. The length of pilot for all the methods is set as $64$. The power ratio between the pilot and the data phase is set as $1$. The error probability threshold is $0.05$. We note that for our proposed method, the channel estimation in the pilot phase is used as the initialization of the TLMP in the data phase. Fig. \ref{ebn01} and Fig. \ref{ebn02} show that our proposed TLMP achieve better energy efficiency compared with all the pilot-based baselines in all the considered settings. Particularly, our TLMP is also better than the FASURA, especially for the case with large $K$, which is the state-of-the-art approach in the massive URA. This is because our proposed TLMP jointly perform the channel estimation and decoding. The channel coefficients are considered as the random variables instead of being considered as fixed constants, and constantly updated based on the observations in the data phase and the newly obtained decoding results.

\begin{figure}[!t]
\centering
\includegraphics[width=3.2in]{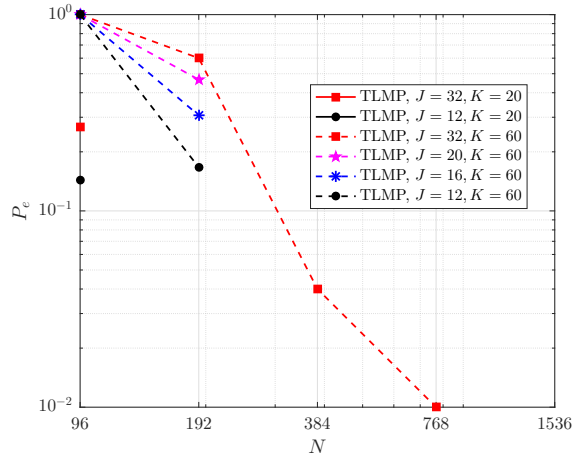}
\caption{Error probability $P_e$ as a function of the length of codewords $N$, the number of blocks $J$ and the number of users $K$.}
\label{sec_size}
\end{figure}

Fig. \ref{sec_size} investigates the impacts of the number of blocks on the error probability performance with $B = 96$ and $M = 100$. The settings of the SNR and channel model are similar with the Figs. \ref{com_coll} and \ref{KM}. We note that since we fix the number of bits of the overall payload, the larger the number of blocks $J$, the shorter of the number of the bits $L$ per block. We can observe from Fig. \ref{sec_size} that decreasing $L$ will cause performance loss of the TLMP, but the TLMP is robust to the short $L$. For different $L$, the spectral efficiency required by TLMP to achieve negligible error probability is very close. Even in the case of $K = 60$ and $J = 32$, i.e., $2^L = 8$, the TLMP also can achieve negligible error probability with $N = 1536$. This observation is a significant distinction between the TLMP and other algorithms in the URA literature, since they always require that $2^L\gg K$ to achieve affordable recovery error. However,  this will cause a large computational complexity in massive MIMO URA. The reason that the TLMP does not require $2^L\gg K$ is that after channel equalization, the problem reduces to $K$ independent SPARCs decoding problems, as seen in (\ref{ftxo}) and (\ref{xtf1}).

\begin{figure}[!t]
\centering
\includegraphics[width=3.2in]{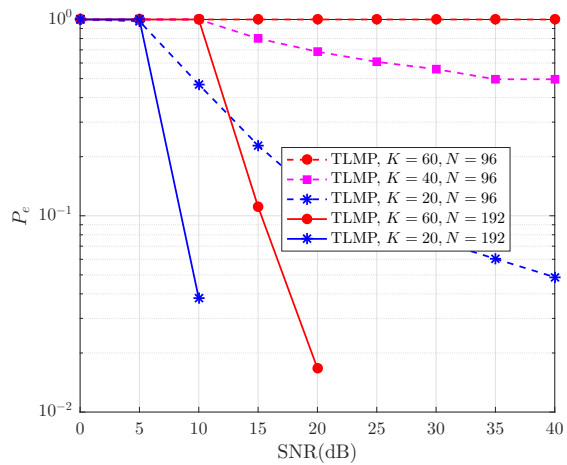}
\caption{Error probability $P_e$ as a function of SNR, the length of codewords $N$ and the number of users $K$.}
\label{power}
\end{figure}

We then demonstrate the error probability of TLMP as a function of SNR, the length of codewords $N$ and the number of users $K$ in Fig. \ref{power}. We set $M = 200$, $L = 8$ and $B = 96$. We can see from Fig. \ref{power} that with a lower $N$, the benefits of increasing the SNR for the error probability will gradually decrease. When the number of users and the length of the codewords gradually approach, the impact of SNR on decoding results is reduced. On the other hand, when providing a relatively large $N$, increasing SNR will bring an apparent performance gain, and the error probability quickly becomes negligible.

\section{Conclusion}
In this paper, we have proposed a novel TLMP algorithm in the massive MIMO URA. Different from the existing works, the payload of each user device was SPARCs coded without redundant parity bits and the decoder jointly recovered all the blocks, acquiring both the information sequences and the massive MIMO channel based on the established Bayesian model. To mitigate the issues that the forms of messages are too complicated, and the total number of messages is too large, we used the core idea of the VMP and AMP. The whole decoding algorithm can be divided into two parts, called the channel equalization part and the decoding part, respectively. In the channel equalization part, the VMP converts the complex problem into a simpler bilinear problem by minimizing the KL divergence, and we therefore integrate the BiG-AMP into this part, while the estimations of the marginal posterior distributions of channel coefficients are updated. In the decoding part, the sub-problem can be tailored into a vector-valued linear problem, and we then propose a Ve-GAMP for the sub-problem, and the SPARCs are decoded based on the derived approximate posterior distributions. We have shown that because of the core idea of jointly recovering, our proposed TLMP significantly enhances the spectral efficiency and energy efficiency compared with the state-of-the-arts baselines, and our algorithm was more robust to the possible codeword collisions.

\appendix
\subsection{BiG-AMP Derivations for Part I of TLMP}
\label{1}
We begin by defining the mean and variance associated with the messages $\Delta_{g_{nm}\leftarrow h_{km}}(h_{km})$ and $\Delta_{g_{nm}\leftarrow r_{nq}}(r_{nq})$, given by
$\hat{h}_{kmn}$, $V_{kmn}^h$ and $\hat{r}_{nqm}$, $V_{nqm}^r$, respectively. Adopting the general assumption, similarly with that in \cite{Kabashima2016}, the prior distribution over each $z_{nm}$ on the condition of $r_{nk}$ can be approximated as a Gaussian distribution.
By defining a function $\mathcal H_{nm}(\cdot)$ corresponds to each $z_{nm}$ as
$
\mathcal{H}_{nm}(\hat q, V_q, y_{nm}) = \log \int{\rm d}z_{nm}p(y_{nm}|z_{nm})\mathcal{N}(z_{nm};\hat q, V_q)
$
and expanding the message $\Delta_{g_{nm}\to r_{nk}}(r_{nk})$ in point $r_{nk} = \hat{r}_{nk}$ and ignoring the $\mathcal{O}(1/N^{3/2})$ term, we can get the approximation
\begin{align}
\Delta_{g_{nm}\to r_{nk}}(r_{nk})&\approx {\rm const}+\big(\hat{h}_{kmn}\mathcal{H}_{nm}'(\hat{p}_{nm}, V_{nm}^p, y_{nm})\nonumber\\
&-\hat{h}_{km}^2\hat{r}_{nk}\mathcal{H}_{nm}''(\hat{p}_{nm}, V_{nm}^p, y_{nm})\big)r_{nk}\nonumber\\
&+\big(\dot{\mathcal{H}}_{nm}(\hat{p}_{nm}, V_{nm}^p, y_{nm})V_{km}^h\nonumber\\
&+\frac{1}{2}\hat{h}_{km}^2\mathcal{H}_{nm}''(\hat{p}_{nm}, V_{nm}^p, y_{nm})\big)r^2_{nk},
\end{align}
where we define variables
\begin{align}
\hat{p}_{nm}&= \sum\nolimits_{k}\hat{r}_{nkm}\hat{h}_{kmn},\nonumber\\
 V_{nm}^p&= \sum\nolimits_{k}(V_{kmn}^hr^2_{nkm}+V_{nkm}^r\hat{h}^2_{kmn}+V_{kmn}^hV_{nkm}^r).\label{vp}
\end{align}
$\mathcal{H}_{nm}'(\cdot)$ and $\mathcal{H}_{nm}''(\cdot)$ are the first and second order derivatives of the function $\mathcal{H}_{nm}(\cdot)$ with respect to the first argument, while $\dot{\mathcal{H}}_{nm}(\cdot)$ is the first derivative with respect to the second argument. These derivatives have the following relation:
\begin{align}
&\dot{\mathcal{H}}_{nm}(\hat{p}_{nm}, V_{nm}^p, y_{nm})\nonumber\\
 &= \frac{1}{2}\bigg(\big(\mathcal{H}_{nm}'(\hat{p}_{nm}, V_{nm}^p, y_{nm})\big)^2+\mathcal{H}_{nm}''(\hat{p}_{nm}, V_{nm}^p, y_{nm})\bigg).\nonumber
\end{align}
We further define a pair of variables, i.e., scaled residual $\hat{s}_{nm}$ and inverse scaled variance $V_{nm}^s$. Using the basic theory of the exponential family \cite{GAMP}, we have
\begin{align}
\hat{s}_{nm} &= \mathcal{H}_{nm}'(\hat{p}_{nm}, V_{nm}^p, y_{nm})= \frac{1}{V_{nm}^p}(\hat{z}_{nm}-\hat{p}_{nm}),\\
V_{nm}^s &= -\mathcal{H}_{nm}''(\hat{p}_{nm}, V_{nm}^p, y_{nm}) = \frac{1}{V_{nm}^p}\left(1-\frac{V_{nm}^z}{V_{nm}^p}\right),
\end{align}
where $\hat{z}_{nm} = \langle{z}_{nm};\hat{p}_{nm}, V_{nm}^p, y_{nm}\rangle$, and $V_{nm}^z = {\rm Var}({z}_{nm};\hat{p}_{nm}, V_{nm}^p, y_{nm})$. The mean and variance are taken over the distribution
$
p(z_{nm};\hat{p}_{nm}, V_{nm}^p, y_{nm}) = \frac{1}{C}p(y_{nm}|z_{nm})\mathcal{N}(z_{nm};\hat{p}_{nm}, V_{nm}^p)
$,
where $C$ is a normalized constant. Based on (\ref{like}), we therefore have \begin{align}
\hat{z}_{nm} = \frac{y_{nm}V_{nm}^p+\hat{p}_{nm}\sigma^2}{V_{nm}^p+\sigma^2},\quad
V_{nm}^z = \frac{V_{nm}^p\sigma^2}{V_{nm}^p+\sigma^2}.
\end{align}
Consequently, the message propagated from node $r_{nk}$ to node $g_{nm}$ can be formulated by
\begin{align}
&\Delta_{g_{nm}\to r_{nk}}(r_{nk})\approx {\rm const}+(\hat{h}_{kmn}\hat{s}_{nm}+V_{nm}^s\hat{h}_{km}^2\hat{r}_{nk})r_{nk}\nonumber\\
&-\frac{1}{2}\big(V_{nm}^s\hat{h}_{km}^2-V_{km}^h(\hat{s}_{nm}^2-V_{nm}^s)\big)r_{nk}^2.\label{gtr}
\end{align}
Similarly, we also have
\begin{align}
&\Delta_{g_{nm}\to h_{km}}(h_{km})\approx {\rm const}+(\hat{r}_{nkm}\hat{s}_{nm}+V_{nm}^s\hat{r}_{nk}^2\hat{h}_{km})h_{km}\nonumber\\
&-\frac{1}{2}\big(V_{nm}^s\hat{r}_{nk}^2-V_{nk}^r(\hat{s}_{nm}^2-V_{nm}^s)\big)h_{km}^2.\label{gth}
\end{align}

We note that both the messages in (\ref{gtr}) and (\ref{gth}) are Gaussian. After that, we then derive the message propagated from the variable nodes to the factor nodes. For the message $\Delta_{g_{nm}\leftarrow r_{nk}(r_{nk})}$, we define
\begin{align}
V_{nk}^u &= \big(\sum\nolimits_{m}V_{nm}^s\hat{h}_{km}^2-V_{km}^h(\hat{s}_{nm}^2-V_{nm}^s)\big)^{-1},\nonumber\\
\hat{u}_{nk} &= \hat{r}_{nk}\big(1+V_{nk}^u\sum\nolimits_{m}V_{km}^h(\hat{s}_{nk}^2-V_{nk}^s)\big)\nonumber\\
&\quad\quad+V_{nk}^u\sum\nolimits_{m}\hat{s}_{nm}\hat{h}_{kmn},\label{hatu}
\end{align}
Based on (\ref{gtr}), the expectation of (\ref{rtg1}) can be approximated as $
\hat{r}_{nkm}\approx \hat{r}_{nk}-\hat{s}_{nm}\hat{h}_{km}V_{nk}^r$,
where $\hat{r}_{nk}$ and $V_{nk}^r$ can be regarded as the mean and variance of PDF $\frac{1}{C}\exp\big(\Delta_{r_{nk}\leftarrow f_{nk}}(r_{nk})\big)\mathcal{N}(r_{nk};\hat{u}_{nk}, V_{nk}^u)$.
Similarly, we define
\begin{align}
V_{km}^w &= \big(\sum\nolimits_{n}V_{nm}^s\hat{r}_{nk}^2-V_{nk}^{r}(\hat{s}_{nm}^2-V_{nm}^s)\big)^{-1},\nonumber\\
\hat{w}_{km} &= \hat{h}_{km}\big(1+V_{km}^w\sum\nolimits_{n}V_{nk}^r(\hat{s}_{nm}^2-V_{nm}^s)\big)\nonumber\\
&\quad\quad+V_{km}^w\sum\nolimits_{n}\hat{s}_{nm}\hat{r}_{nkm},\label{hatw}
\end{align}
and we can obtain $\hat{h}_{kmn} \approx \hat{h}_{km}-\hat{s}_{nm}\hat{r}_{nk}V_{km}^h$, where $\hat{h}_{km}$ and $V_{km}^h$ are the mean and variance of the PDF $\frac{1}{C}\tilde{p}(h_{km})\mathcal{N}(h_{km};\hat{w}_{km}, V_{km}^w)$. Dropping the term of $\mathcal{O}(1/N^{1/2})$, we get
\begin{align}
\hat{p}_{nm}\approx \bar{p}_{nm}-\hat{s}_{nm}\bar{V}_{nm}^p,\quad V_{nm}^p \approx \bar{V}_{nm}^p+\sum\nolimits_{k}V_{km}^hV_{nk}^r,\nonumber
\end{align}
where
$
\bar{p}_{nm} = \sum\limits_{k}\hat{r}_{nk}\hat{h}_{km}$ and $\bar{V}_{nm}^p= \sum\limits_{k}V_{km}^hr^2_{nk}+V_{nk}^r\hat{h}^2_{km}
$. Accordingly, we have
\begin{align}
\hat{u}_{nk}&\approx \hat{r}_{nk}(1-V_{nk}^u\sum\nolimits_{m}V_{km}^hV_{nm}^s)+V_{nk}^u\sum\nolimits_{m}\hat s_{nm}\hat h_{km},\nonumber\\
\hat{w}_{km}&\approx \hat{h}_{km}(1-V_{km}^w\sum\nolimits_{n}V_{nk}^rV_{nm}^s)+V_{km}^w\sum\nolimits_{n}\hat s_{nm}\hat r_{nk},\nonumber
\end{align}
and based on the clarification in \cite{big-amp}, the following approximations hold.
\begin{align}
V_{nk}^u\approx \big(\sum\nolimits_{m}V_{nm}^s\hat{h}^2_{km}\big)^{-1},\quad V_{km}^w\approx \big(\sum\nolimits_{n}V_{nm}^s\hat{r}^2_{nk}\big)^{-1}.\nonumber
\end{align}
\subsection{Ve-GAMP Derivations for Part II of TLMP}\label{2}
We define the mean and covariance of message $\exp(\Delta_{f_{nk}\leftarrow \mathbf x_{jk}}(\mathbf x_{jk}))$ as $\hat{\mathbf x}_{jkn}$ and $\mathbf C_{jkn}^{x}$, respectively. By the general Gaussian assumption, the prior distribution over each the mixing variable $r_{nk}$ can be approximated as a Gaussian distribution, whose mean and variance can be easily calculated by the defined statistics, given by
\begin{align}
p(r_{nk}|\mathbf{x}_{jk})\approx \mathcal{N}(r_{nk};\sum\limits_{p\neq j}\mathbf a^T_{np}\hat{\mathbf x}_{pkn}+\mathbf a^T_{nj}\mathbf{x}_{jk}, \sum\limits_{p\neq j}\mathbf{a}_{np}^T\mathbf C_{pkn}^{x}\mathbf{a}_{np}).\nonumber
\end{align}
We then define a function
\begin{align}
\mathcal{T}_{nk}(\hat{q},V^q) &= \int\log{\rm d}r_{nk}\exp\big(\Delta_{r_{nk}\rightarrow f_{nk}}(r_{nk})\big)\nonumber\\
&\times\mathcal{N}(r_{nk};\hat{q},V^q).\label{fun_T}
\end{align}
and some useful variables
\begin{align}
\hat{q}_{nk} =  \sum\nolimits_{j}\mathbf a^T_{nj}\hat{\mathbf x}_{jkn},\quad V_{nk}^q  = \sum\nolimits_{j}\mathbf{a}_{nj}^T\mathbf C_{jkn}^{x}\mathbf{a}_{nj}.\label{hatq}
\end{align}
Combining (\ref{fun_T}) and (\ref{hatq}) with (\ref{ftxo}), and expanding the message $\Delta_{f_{nk}\rightarrow \mathbf{x}_{jk}}(\mathbf{x}_{jk})$ in point $\mathbf{x}_{jk} = \hat{\mathbf{x}}_{jk}$, we can obtain the following message by ignoring the $\mathcal{O}(1/N^{3/2})$ term as
\begin{align}
&\Delta_{f_{nk}\rightarrow \mathbf{x}_{jk}}(\mathbf{x}_{jk})\approx {\rm const}+\frac{1}{2}\mathcal{T}_{nk}''(\hat{q}_{nk},V^q_{nk})\hat{\mathbf{x}}_{jk}^T\mathbf{a}_{nj}\mathbf{a}_{nj}^T\hat{\mathbf{x}}_{jk}\nonumber\\
&+\big(\mathcal{T}_{nk}'(\hat{q}_{nk},V^q_{nk})\mathbf{a}_{nj}^T-\hat{\mathbf{x}}_{jk}^T\mathbf{a}_{nj}\mathbf{a}_{nj}^T\mathcal{T}_{nk}''(\hat{q}_{nk},V^q_{nk})\big)\mathbf{x}_{jk}\nonumber,
\end{align}
where $\mathcal{T}_{nk}'(\cdot)$ and $\mathcal{T}_{nk}''(\cdot)$ are the first and second order derivatives of the function $\mathcal{T}_{nk}(\cdot)$ with respect to the first argument. We further define the scale residual $\hat{t}_{nk}$ and the inverse scaled variance $V_{nk}^t$,
\begin{align}
\hat{t}_{nk} &=  \mathcal{T}_{nk}'(\hat{q}_{nk},V^q_{nk}) = {1/V^q_{nk}}(\tilde{r}_{nk}-\hat{q}_{nk}),\nonumber\\
V_{nk}^t &= -\mathcal{T}_{nk}''(\hat{q}_{nk},V^q_{nk}) = {1}/{V^q_{nk}}(1-{\tilde{V}^r_{nk}}/{V^q_{nk}})\nonumber,
\end{align}
where $\tilde{r}_{nk} = \langle{r}_{nk};\hat{q}_{nk},V^q_{nk}\rangle$, $\tilde{V}^r_{nk} = {\rm Var}({r}_{nk};\hat{q}_{nk},V^q_{nk})$, and the moments are taken over the distribution
$
p({r}_{nk};\hat{q}_{nk},V^q_{nk}) = C^{-1}\exp\big(\Delta_{r_{nk}\rightarrow f_{nk}}(r_{nk})\big)\mathcal{N}(r_{nk};\hat{q}_{nk},V^q_{nk}).
$
As a consequence, we get
\begin{align}
&\Delta_{f_{nk}\rightarrow \mathbf{x}_{jk}}(\mathbf{x}_{jk})\approx {\rm const}+(\hat{t}_{nk}\mathbf{a}_{nj}^T+\hat{\mathbf{x}}_{jk}^T\mathbf{a}_{nj}\mathbf{a}_{nj}^TV_{nk}^t)\mathbf{x}_{jk}\nonumber\\
&+\frac{1}{2}V_{nk}^t\hat{\mathbf{x}}_{jk}^T\mathbf{a}_{nj}\mathbf{a}_{nj}^T\hat{\mathbf{x}}_{jk}.\label{ftx}
\end{align}

Based on the (\ref{ftx}), the message (\ref{xtf1}) from the variable node $\mathbf x_{jk}$ to the factor node $f_{nk}$ can be formulated as
$
\Delta_{f_{nk}\leftarrow \mathbf{x}_{jk}}(\mathbf{x}_{jk}) \approx \log p(\mathbf{x}_{jk})-\frac{1}{2}(\mathbf{x}_{jk}-\mathbf{\hat{d}}_{jkn})^T\mathbf C_{jkn}^{d-1}(\mathbf{x}_{jk}-\mathbf{\hat{d}}_{jkn})
$
, where we define
\begin{align}
\mathbf C_{jkn}^{d}  = \big(\sum\limits_{v\neq n} V_{vk}^t\mathbf{a}_{vj}\mathbf{a}_{vj}^T\big)^{-1},\hat{\mathbf d}_{jkn} = \mathbf C_{jkn}^{d}\sum\limits_{v\neq n}\hat{t}_{vk}\mathbf{a}_{vj}+\hat{\mathbf{x}}_{jk}.\nonumber
\end{align}
We then define
\begin{align}
 \hat{\mathbf x}_{jkn} = \boldsymbol{\mathcal L}(\hat{\mathbf d}_{jkn}, \mathbf C_{jkn}^{d})=\frac{\int{\rm d}\mathbf x_{jk}\mathbf x_{jk}p(\mathbf x_{jk})\mathcal{N}(\mathbf x_{jk}; \hat{\mathbf d}_{jkn}, \mathbf C_{jkn}^{d})}{\mathcal{A}(\hat{\mathbf d}_{jkn}, \mathbf C_{jkn}^{d})},\nonumber
\end{align}
where $\mathcal{A}(\hat{\mathbf d}_{jkn}, \mathbf C_{jkn}^{d})=\int{\rm d}\mathbf x_{jk}p(\mathbf x_{jk})\mathcal{N}(\mathbf x_{jk}; \hat{\mathbf d}_{jkn}, \mathbf C_{jkn}^{d})$. The function $\mathcal{A}(\hat{\mathbf d}_{jkn}, \mathbf C_{jkn}^{d})$ is exactly the partition function over the distribution
\begin{align}
&p(\mathbf x_{jk}; \hat{\mathbf d}_{jkn}, \mathbf C_{jkn}^{d})\nonumber\\
 =& \exp\big({\Phi}_{jkn}(\mathbf x_{jk};\mathbf C_{jkn}^{d})-\frac{1}{2}\hat{\mathbf d}_{jkn}^T\mathbf C_{jkn}^{d-1}\hat{\mathbf d}_{jkn}+\hat{\mathbf d}_{jkn}^T\mathbf C_{jkn}^{d-1}\mathbf x_{jk}\big) \nonumber\\
=& \frac{1}{C}p(\mathbf x_{jk})\mathcal{N}(\mathbf x_{jk}; \hat{\mathbf d}_{jkn}, \mathbf C_{jkn}^{d}).\label{dis_x}
\end{align}
We immediately have
\begin{align}
&\frac{\partial\mathcal{A}(\hat{\mathbf d}_{jkn}, \mathbf C_{jkn}^{d})}{\partial\hat{\mathbf d}_{jkn}} = \mathbf C_{jkn}^{d-1}\langle\mathbf x_{jk};\hat{\mathbf d}_{jkn}, \mathbf C_{jkn}^{d}\rangle,\nonumber\\
&\frac{\partial^2\mathcal{A}(\hat{\mathbf d}_{jkn}, \mathbf C_{jkn}^{d})}{\partial\hat{\mathbf d}_{jkn}\partial\hat{\mathbf d}_{jkn}^T} = \mathbf C_{jkn}^{d-1}{\rm Var}(\mathbf x_{jk};\hat{\mathbf d}_{jkn}, \mathbf C_{jkn}^{d})\mathbf C_{jkn}^{d-1},\nonumber
\end{align}
where $\langle\mathbf x_{jk};\hat{\mathbf d}_{jkn}, \mathbf C_{jkn}^{d}\rangle$ and ${\rm Var}(\mathbf x_{jk};\hat{\mathbf d}_{jkn}, \mathbf C_{jkn}^{d})$ are the mean and covariance matrix over distribution (\ref{dis_x}). Obviously, we have
$
\mathbf C_{jkn}^{x} = {\partial\boldsymbol{\mathcal L}(\hat{\mathbf d}_{jkn}, \mathbf C_{jkn}^{d})}/{\partial\hat{\mathbf d}_{jkn}^T}\mathbf C_{jkn}^{d}$. We further define the variable $\mathbf C_{jk}^{d}= \big(\sum\nolimits_{n}V_{nk}^t\mathbf{a}_{nj}\mathbf{a}_{nj}^T\big)^{-1}$ and $\hat{\mathbf d}_{jk} = \mathbf C_{jk}^{d}\sum\limits_{n}\hat{t}_{nk}\mathbf{a}_{nj}+\hat{\mathbf{x}}_{jk}$, together with the matrix inversion lemma, we have
\begin{align}
&\mathbf C_{jkn}^{d}-\mathbf C_{jk}^{d}= \boldsymbol{\mathcal{O}}(1/N^2)\nonumber\\
 &= \big(\sum\nolimits_{n}V_{nk}^t\mathbf{a}_{nj}\mathbf{a}_{nj}^T\big)^{-1}V_{nk}^t\mathbf{a}_{nj}\mathbf{a}_{nj}^T\big(\sum\nolimits_{v\neq n}V_{vk}^t\mathbf{a}_{vj}\mathbf{a}_{vj}^T\big)^{-1},\nonumber\\
&\hat{\mathbf d}_{jkn}-\hat{\mathbf d}_{jk} = -\mathbf C_{jk}^{d}\hat{t}_{nk}\mathbf{a}_{nj}+\boldsymbol{\mathcal{O}}(1/N^{3/2}),\nonumber
\end{align}
where $\boldsymbol{\mathcal{O}}(\cdot)$ is a $L\times L$ matrix with each element equaling ${\mathcal{O}}(\cdot)$. By ignoring the term $\boldsymbol{\mathcal{O}}(1/N^{3/2})$, we get the approximation
$
 \hat{\mathbf x}_{jkn}\approx  \hat{\mathbf x}_{jk}-\mathbf C_{jk}^{x}\hat{t}_{nk}\mathbf{a}_{nj}
$,
where
$
 \hat{\mathbf x}_{jk} = \boldsymbol{\mathcal L}(\hat{\mathbf d}_{jk}, \mathbf C_{jk}^{d})$, $\mathbf C_{jk}^{x} = {\partial\boldsymbol{\mathcal L}(\hat{\mathbf d}_{jk}, \mathbf C_{jk}^{d})}/{\partial\hat{\mathbf d}_{jk}^T}\mathbf C_{jk}^{d}
$, and we therefore obtain
$
\hat{q}_{nk} = \bar{q}_{nk}-\hat{t}_{nk}\bar{V}_{nk}^q
$,
where $
\bar{q}_{nk}  = \sum\nolimits_{j}\mathbf a^T_{nj}\hat{\mathbf x}_{jk}$ and $\bar{V}_{nk}^q  = \sum\nolimits_{j}\mathbf{a}_{nj}^T\mathbf C_{jk}^{x}\mathbf{a}_{nj}$.
Note that we approximately have ${V}_{nk}^q\approx \bar{V}_{nk}^q$ by neglecting the higher order infinite small term.

\bibliography{a}

\end{document}